%% file: paper.tex
\documentclass{CSML}

\def\dOi{11(4:19)2015}
\lmcsheading%
{\dOi}
{1--34}
{}
{}
{Dec.~\phantom02, 2014}
{Dec.~29, 2015}
{}


\ACMCCS{[{\bf Theory of computation}]: Logic---Logic and verification; [{\bf
    Security and privacy}]: Network security---Security protocols}

\usepackage{graphicx} 
\usepackage{xspace, ifthen}
\usepackage{mathpartir} 
\usepackage{amsmath,amssymb} 
\usepackage{color}
\usepackage{multirow}
\usepackage{alltt}
\usepackage{url}
\usepackage{relsize}
\usepackage{algorithm,algpseudocode,mathabx}
\usepackage{hyperref}

\usepackage{verbatim}
\usepackage{pifont}

\input{macros}

\begin{document}
\title[A Program Logic for Verifying Secure Routing Protocols]{A Program Logic
for Verifying\\Secure Routing Protocols}

\author[C.~Chen]{Chen Chen\rsuper a}
\address{{\lsuper{a,c,d}}University of Pennsylvania}
\email{\{chenche,haoxu,boonloo\}@cis.upenn.edu}

\author[L.~Jia]{Limin Jia\rsuper b}
\address{{\lsuper b}Carnegie Mellon University}
\email{liminjia@cmu.edu}

\author[H.~Xu]{Hao Xu\rsuper c}
\address{\vspace{-18 pt}}

\author[C.~Luo]{Cheng Luo\rsuper d}
\address{\vspace{-18 pt}}

\author[W.~Zhou]{Wenchao Zhou\rsuper e}
\address{{\lsuper e}Georgetown University}
\email{wzhou@cs.georgetown.edu}

\author[B.~T.~Loo]{Boon Thau Loo\rsuper f}
\address{\vspace{-18 pt}}

\begin{abstract}

  The Internet, as it stands today, is highly vulnerable to
  attacks. 
  However, little has been done to understand and verify the formal
  security guarantees of proposed secure inter-domain routing
  protocols, such as Secure BGP (S-BGP).  In this paper, we develop a
  sound program logic for \Langsec---a declarative specification
  language for secure routing protocols---for verifying 
  properties of these protocols.  We 
  prove invariant properties of \Langsec programs that run in an
  adversarial environment.  As a step towards automated verification,
  we implement a verification condition generator (VCGen) to
  automatically extract proof obligations.  VCGen is integrated
  into a compiler for \Langsec that can generate executable protocol
  implementations; and thus, both verification and empirical
  evaluation of secure routing protocols can be carried out in this
  unified framework.  To validate our framework, we encoded
  several proposed secure routing mechanisms in \Langsec, verified variants of
  path authenticity properties by manually discharging the generated
  verification conditions in Coq, and generated executable code
  based on \Langsec specification and ran the code in simulation.
\end{abstract}

\maketitle

\section{Introduction}
\label{sec:intro}
In recent years, we have witnessed an explosion of services provided
over the Internet.
These services are increasingly transferring
customers' private information over the network and
used in mission-critical tasks. Central to ensuring the reliability and
security of these services is a secure and efficient Internet routing
infrastructure.
Unfortunately, the Internet infrastructure, as it stands today, is
highly vulnerable to attacks. 
The Internet runs the {\em Border Gateway Protocol} (BGP), where routers are
grouped into Autonomous Systems (\textit{AS}) administrated by
Internet Service Providers (\textit{ISP}s). Individual ASes exchange
route advertisements with neighboring ASes using the {\em path-vector}
protocol. Each originating AS first sends a route advertisement
(containing a single AS number) for the IP prefixes it
owns. Whenever an AS receives a route advertisement, it adds
itself to the AS {\em path}, and advertises the best route to its neighbors
based on its routing policies.  Since these route advertisements
are not authenticated, ASes can advertise non-existent routes
or claim to own IP prefixes that they do not. 

These faults may lead to 
 long periods of interruption of the
Internet; best epitomized by
recent high-profile attacks~\cite{youtube,congressreport}. 
In response to these vulnerabilities, several new
Internet routing architectures and protocols for a more secure
Internet have been proposed.  These range
from security extensions of BGP (Secure-BGP
(S-BGP)~\cite{securebgp}, ps-BGP~\cite{psbgp}, so-BGP~\cite{sobgp}),
to ``clean-slate'' Internet architectural redesigns such as
SCION~\cite{scion} and ICING~\cite{icing}.  However, {\em none} of
the proposals formally analyzed
their security properties. These protocols are implemented
from scratch, evaluated primarily experimentally, and
their security properties shown via informal reasoning.

Existing protocol analysis
tools~\cite{proverif-tool,datta07:entcs,maude-npa} are rarely used in
analyzing routing protocols because they are considerably
more complicated than cryptographic protocols: they often compute
local states, are recursive, and their security properties need to
be shown to hold on arbitrary network topologies. 
 As the number of
models is infinite, model-checking-based tools, in general, cannot be
used to prove the protocol secure. 

 \begin{figure}[t!]
\begin{minipage}[c]{3.1in}
 \centering \includegraphics[scale=0.72]{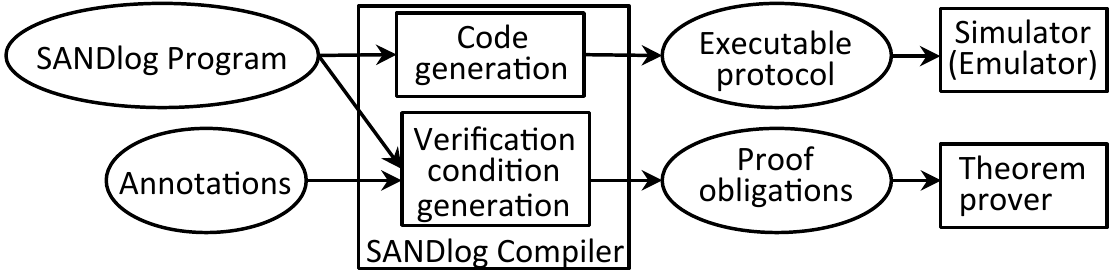}
\end{minipage}
\hfill
\begin{minipage}[c]{1.35in}
{\scriptsize The round objects are code
   (proofs), which are the input or output of the framework. The
   rectangular objects are software components of the framework.}
\end{minipage}

  \caption{\scriptsize Architecture of a unified framework for implementing and
   verifying secure routing protocols.}
  \label{fig:arch}
 \end{figure}

 To overcome the above limitations, we explore a novel proof
 methodology to verify these protocols.  We augment prior work on
 declarative networking (\Lang)~\cite{cacm09} with cryptographic
 libraries to provide compact encoding of secure routing protocols. We
 call our language \Langsec (stands for {\em Secure and Authenticated
   Network DataLog}).  
We develop a program logic for reasoning about \Langsec
programs that execute in an adversarial environment. The properties
proved on a \Langsec program hold even when the program interact with
potentially malicious programs in the network. 

Based on the program logic, we implement a verification condition
generator (\VCG), which takes as inputs the \Langsec program and
user-provided annotations, and outputs intermediary proof obligations
as a Coq file, where proof can be filled. \VCG is integrated into
the \Langsec compiler, an cryptography-augmented extension to the
declarative networking engine RapidNet~\cite{rapidneturl}. 
The compiler
is able to translate our \Langsec specification into executable code,
which is amenable to implementation and evaluation. 

We choose to use a declarative language as our specification language
for two reasons. First, it has been shown that declarative languages
such as \Lang can specify a variety of network protocols
concisely~\cite{cacm09}. Second, \Langsec is the specification
language for both verification and generating low-level
implementations. As a result, 
 verification and empirical evaluation
of secure routing protocols can be carried out in a unified
framework (Figure~\ref{fig:arch}).

We summarize our technical contributions:
\begin{packedenumerate}
\item We define a program logic for verifying \Langsec programs in the presence of adversaries
  (Section~\ref{sec:verification}).  We prove that our logic is sound.
\item We implement \VCG for automatically generating proof
  obligations and integrate \VCG into a compiler for \Langsec (Section~\ref{sec:implementation}).
\item We encode S-BGP and SCION in \Langsec, verify path
authenticity properties of these protocols, and run them in simulation (Section~\ref{sec:casestudy}).
\end{packedenumerate}

\noindent Compared to our conference paper~\cite{forte2014} in FORTE 2014, we have added
the new case study of SCION, a clean-slate Internet architecture for inter-domain
routing. We encode SCION in \Langsec, simulate the code in RapidNet, and
verify variants of route authenticity properties. We also provide
a comparison between S-BGP and SCION. It shows that SCION's security
guarantee in routing is similar to S-BGP, as they both use layered signatures
to protect advertised path from being tampered with by an
attacker. SCION, however, enforces stronger
security properties during data forwarding, enabling an AS to authenticate
an upstream neighbor. On the other hand, S-BGP does not provide any guarantee
regarding data forwarding, which means an AS could forward packets coming from
any neighbor. 
\Langsec specification and formal verification of both solutions can be found online
(\url{http://netdb.cis.upenn.edu/secure_routing/}.)

\section{\Langsec}
\label{sec:sendlog}
We specify secure routing protocols in a distributed declarative programming
language called \Langsec. \Langsec is an extension to Network Datalog
(\Lang)~\cite{cacm09}, which is
proved to be a compact and clean way of specifying network routing
protocols~\cite{declareRoute}. \Langsec inherits the expressiveness of \Lang,
and is augmented with
security primitives (e.g. asymmetric encryption) necessary for specifying secure
routing protocols. 

\subsection{Syntax}
\label{sec:sandlog-syntax}
\Langsec's syntax is summarized in Figure~\ref{fig:ndlog-syntax}.  A
typical \Langsec program is composed of a set of rules, each of which
consists of a rule head and a rule body.
The rule head is a predicate, or tuple (we use predicate and tuple
interchangeably).  A rule body consists of a list of body
elements which are either tuples or atoms (i.e. assignments and
inequality constraints). The head tuple supports aggregation functions
as its arguments, whose semantics will be introduced in
Section~\ref{sec:sendlog:op-semantics}.  \Langsec also defines (and
implements) a number of cryptographic functions, which represent
common encryption operations such as signature generation and
verification.  Intuitively, a \Langsec rule specifies that the head
tuple is derivable if all the body tuples are derivable and all the
constraints represented by the body atoms are satisfied.
\Langsec distinguishes between base tuples and derived
  tuples. Base tuples are populated upon system initialization. Rules
  for populating base tuples are denoted as $b$.

\begin{figure}[t!]
\centering
\[
\begin{array}{l@{~}lcl@{~~}l@{~}lcl}

\textit{Crypt func}& f_c&\bnfdef & \mathsf{f\_sign\_asym} \bnfalt
\mathsf{f\_verify\_asym}\cdots
\\ 
\textit{Atom} & a & \bnfdef  & x := t \bnfalt t_1 \bop t_2
\\
 \textit{Terms} & t & \bnfdef & x \bnfalt c  \bnfalt \nodeid\bnfalt f(\,\vec{t}\,)
 \bnfalt f_c(\,\vec{t}\,)
\\
\textit{Predicate} & \predicate & \bnfdef & p(\halist) 
 \bnfalt p(\balist)
\\
\textit{Body Elem} & B & \bnfdef & p(\balist) 
 \bnfalt a
\\
\textit{Arg List} & \alist & \bnfdef &\cdot \bnfalt \alist, x\bnfalt
\alist, c
\\
\textit{Rule Body} & \rbody & \bnfdef & \cdot \bnfalt \rbody, B
\\
\textit{Body Args} & \balist & \bnfdef & @\nodeid, \alist
\\
\textit{Rule} & \rules & \bnfdef & p(\halist) \derives \rbody
\\
\textit{Head Args} & \halist & \bnfdef & \balist \bnfalt @\nodeid, \alist,
\agf\langle x \rangle, \alist
\\
\textit{Base tp rules} & b & \bnfdef & p(\halist).
\\
\textit{Program} & \prog(\nodeid) & \bnfdef & b_1,\cdots, b_n, \rules_1, \cdots, \rules_k 

\end{array}
\]
 \caption{Syntax of \Langsec}
 \label{fig:ndlog-syntax}
 \end{figure}

To support distributed execution, 
a \Langsec program \prog is
parametrized over the node it runs on. Each tuple in the program is supposed to
have a location
specifier, written $@\nodeid$, which specifies where a tuple resides and 
serves as the first argument of a tuple. A rule head can specify a location different from its body
tuples. When such a rule is executed, the
derived tuple is sent to the remote node represented by the location
specifier of the head tuple. We discuss the operational semantics of
\Langsec in detail in Section~\ref{sec:sendlog:op-semantics}.

To specify security operations in secure routing protocols, our syntax
definition also includes cryptographic functions.
Figure~\ref{fig:cryp} gives detailed explanation of these functions. 
Users can add additional cryptographic primitives to \Langsec based on their needs.

\begin{figure}[b!]
\centering
  \begin{tabular}{| l | l |}
    \hline
    \multicolumn{1}{|c|}{\textbf{Function}} & 
    \multicolumn{1}{c|}{\textbf{Description}} \\ \hline
    f\_sign\_asym(\textit{info}, \textit{key}) & 
    Create a signature of \textit{info} using \textit{key} \\ \hline

    f\_verify\_asym(\textit{info}, \textit{sig}, \textit{key}) &
    Verify that \textit{sig} is the signature of \textit{info} using
    \textit{key} \\ \hline
    f\_mac(\textit{info}, \textit{key}) &
    Create a message authentication code of \textit{info}
    using \textit{key} \\ \hline
    f\_verifymac(\textit{info}, \textit{MAC}, \textit{key}) &
    Verify \textit{info} against \textit{MAC} using \textit{key}
    \\ \hline
  \end{tabular}
  \caption{Cryptographic functions in \Langsec}
  \label{fig:cryp}
\end{figure}

\Paragraph{An example program.} 
In Figure~\ref{fig:bestpath}, we show an example program for computing the
shortest path between each pair of nodes in a network. $s$ is the
location parameter of the program, representing the ID of the node
where the program is executing. Each node stores three kinds of
tuples: $\pred{link}(@s, d, c)$ means that there is a direct link from
$s$ to $d$ with cost $c$; $\pred{path}(@s, d, c, p)$ means that $p$ is
a path from $s$ to $d$ with cost $c$; and $\pred{bestPath}(@s, d, c,
p)$ states that $p$ is the lowest-cost path between $s$ and $d$.
Here, \pred{link} is a base tuple, whose values are determined by the
  concrete network topology. $\pred{path}$ and $\pred{bestPath}$ are derived tuples. Figure~\ref{fig:bestpath} only shows the rules
  common to all network nodes. Rules for initializing the base tuple
  $\pred{link}$ depend on the topology and are omitted from the
  figure.

\begin{figure}[t!]
\centering
\[
\hspace{-10pt}\small
\begin{array}{l@{~~}l}
\textit{sp1}&\pred{path}(@s, d, c, p) \derives \pred{link}(@s, d, c), p:=[s,d].
\\
\textit{sp2}&\pred{path}(@z, d, c, p) \derives \pred{link}(@s, z, c1), \pred{path}(@s, d, c2, p1), 
c := c1 + c2, p := z{::}p1. 
\\
\textit{sp3}&\pred{bestPath}(@s, d, \pred{min}\langle c \rangle, p) \derives \pred{path}(@s, d, c, p).
\end{array}
\]
 \caption{A \Langsec program for computing all-pair shortest paths}
 \label{fig:bestpath}
\end{figure}

In the program, rule {\em sp1} computes all one-hop paths based on direct links. Rule
{\em sp2} expresses that if there is a link from {\em s} to {\em z} of
cost {\em c1} and a path from {\em s} to {\em d} of cost {\em c2},
then there is a path from {\em z} to {\em d} with cost {\em c1+c2}
(for simplicity, we assume links are symmetric, i.e. if there is a
link from {\em s} to {\em d} with cost {\em c}, then a link from {\em
  d} to {\em s} with the same cost {\em c} also exists). Finally, rule
{\em sp3} aggregates all paths with the same pair of source and
destination ($s$ and $d$) to compute the shortest path. The arguments
that appear before the aggregation denotes the group-by keys.

 We can construct a more secure variant of the
 shortest path protocol by deploying
 signature authentication in 
 the rules involving cross-node communications (e.g. $\f{sp2}$).
 In the following rule $\f{sp2'}$, a signature $\f{sig}$ for the path
 becomes an additional argument to the \pred{path} tuple.
 When node $s$ receives such a tuple, it verifies the
 signature of the path
 $\pred{f\_verify}(p1, \f{sig}, \f{pk})$. 
 When $s$ sends out a path to its neighbor, it generates a signature
 by assigning $\f{sig} := \pred{f\_sign}(p, \f{sk})$. Here \pred{f\_sign} and
 \pred{f\_verify} are user-defined asymmetric cryptographic functions (e.g. RSA).

 \vspace{3pt}
 \[
 \hspace{-10pt}\small
 \begin{array}{l}
 \f{sp2'}~~\pred{path}(@z, d, c, p, \f{sig}) \derives
 \\ ~~~~~~~~~  \pred{link}(@s, z, c1), 
 \pred{path}(@s, d, c2, p1, \f{sig1}), c := c1 + c2, p := z{::}p1, 
 \\
 ~~~~~~~~~ \pred{pubK}(@s, d, \f{pk}),
 \pred{f\_verify}(p1, \f{sig1}, \f{pk}) = 1,
  \pred{privK}(@s, \f{sk}),
  \f{sig} := \pred{f\_sign}(p, \f{sk}).
 \end{array}
 \]\medskip

\noindent To execute the program, a user provides
   rules for initializing base tuples. For example, if we
   would like to run the shortest-path program over the topology given
   in Figure~\ref{fig:scenario}, the following rules will be
 included in the program. Rules $\f{rb}1$ lives at node $A$, rules
 $\f{rb}2$ and $\f{rb}3$ live at node $B$, and rule $\f{rb}4$ lives at
node $C$.

 \vspace{3pt}
 \[\hspace{-10pt}\small
 \begin{array}{lr}
 \f{rb}1~~\pred{link}(@A, B, 1). & 
 ~~~~~~~~~~~~~\f{rb}3~~\pred{link}(@B, C, 1).
 \\
 \f{rb}2~~\pred{link}(@B, A, 1). & 
 ~~~~~~~~~~~~~\f{rb}4~~\pred{link}(@C, B, 1).
 \end{array}
 \]

\subsection{Operational Semantics}
\label{sec:sendlog:op-semantics}
The operational semantics of \Langsec adopts a distributed state
transition model. 
Each node runs a designated \Langsec program, and maintains a database
of derived tuples as its local state. Nodes can communicate with each
other by sending tuples over the network, which is represented as a
global network queue.  The evaluation of the \Langsec programs follows
the PSN algorithm~\cite{declareNetworks}, and updates the database
incrementally. The semantics introduced here is similar to that of
\Lang, except that we make explicit which tuples are derived, which
are received, and which are sent over the network. This addition is
crucial to specifying and proving protocol properties.

At a high-level, each node computes its local fixed-point by firing
the rules on newly-derived tuples. The fixed-point computation can
also be triggered when a node receives tuples from the network. When a
tuple is derived, it is sent to the node specified by its location
specifier. Instead of blindly computing the fixed-point, we make sure
that only rules whose body tuples are updated are fired. The
operational semantics also support deletion of tuples. A deletion is
propagated through the rules similar to an insertion.

More formally, the constructs needed for defining the operational
semantics of \Langsec are presented below. 
\[\small
\begin{array}{l@{~~}lcl@{\qquad}l@{~~}lcl}
\textit{Table} & \tuples & \bnfdef & \cdot 
  \bnfalt \tuples, (n, P)
&
\textit{Network Queue} & \queue & \bnfdef & \updates
\\
\textit{Update} & \upd & \bnfdef &  \del P \bnfalt 
  \ins P
&
\textit{Local State} & \lstate & \bnfdef & (\nodeid, \tuples,
\updates, \prog(\nodeid))
\\
\textit{Update List} & \updates & \bnfdef &
[\upd_1, \cdots, \upd_n]
&
\textit{Configuration}& \conf & \bnfdef &\queue\rhd \lstate_1, \cdots, \lstate_n
\\
\textit{Trace} & \trace & \bnfdef &
\multicolumn{5}{l}{\steps{\tau_0}\conf_1 \steps{\tau_1} \conf_2 \cdots
  \steps{\tau_n}\conf_{n+1}}
\end{array}
\]
We write $P$ to denote tuples. The database for storing all derived tuples
on a node is denoted \tuples. Because there could be multiple
derivations of the same tuple, we associate each tuple with a
reference count $n$, 
recording the number of valid derivations for
that tuple. 
An update is either an insertion of a tuple, denoted $\ins P$, or
a deletion of a tuple, denoted $\del P$. \reviewer{Might want to mention upfront better what -P is for, and/or
give a forward reference to pg 8}We write \updates to
denote a list of updates. A node's local state, denoted \lstate,
consists of the node's identifier \nodeid, the database \tuples, a
list of unprocessed updates \updates, and the program \prog that
\nodeid runs.  A configuration of the network, written \conf, is
composed of a network update queue \queue, and the set of the local
states of all the nodes in the network. The queue \queue models the
update messages sent across the network. Finally, a trace \trace is a sequence
of time-stamped (i.e. $\tau_i$) configuration transitions. 

\begin{figure}[t]
\centering
\includegraphics[width =\linewidth]{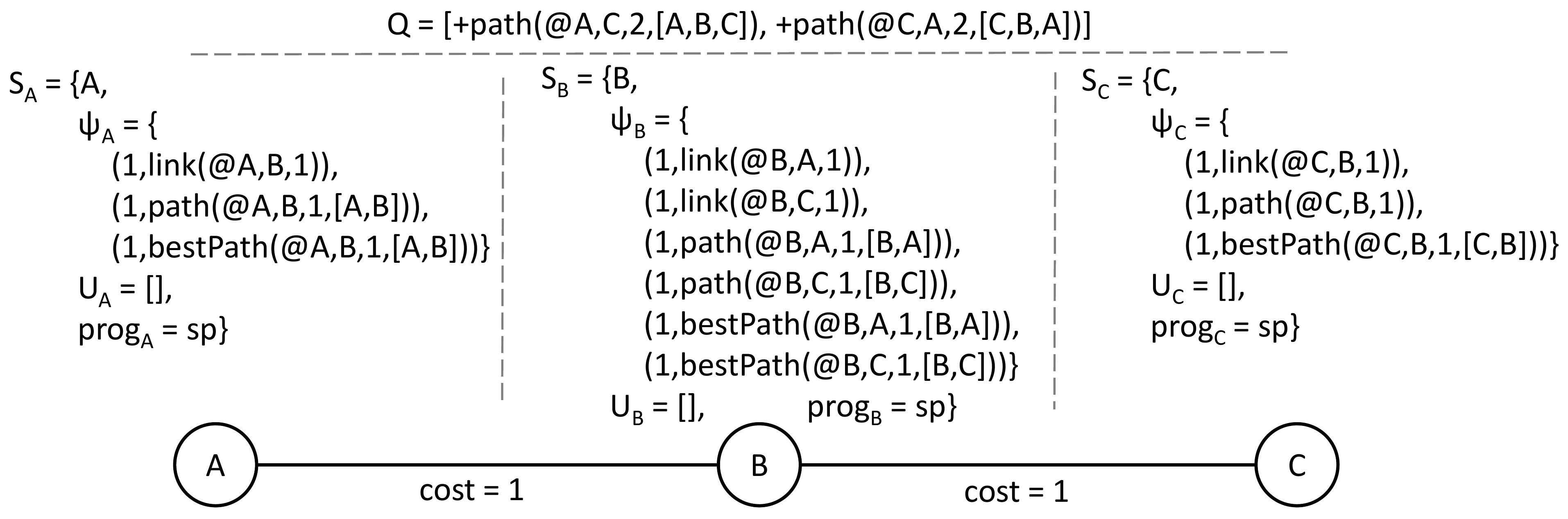}
\caption{An Example Scenario.}\label{fig:scenario}

\end{figure}

Figure~\ref{fig:scenario} presents an example scenario of 
executing the shortest-path program shown in
Section~\ref{sec:sandlog-syntax}.  The network consists of three nodes,
$A$, $B$ and $C$, connected by two links with cost 1. Each node's local state is
displayed right above the node. For example, the local state of the node A is
given by $S_A$ above it. The network queue $Q$ is presented at the top of
Figure~\ref{fig:scenario}. In the
current state, all three nodes are aware of their direct neighbors,
i.e., \pred{link} tuples are in their databases $\tuples_A$,
$\tuples_B$ and $\tuples_C$. They have constructed paths to their
neighbors (i.e., the corresponding \pred{path} and \pred{bestPath} tuples are stored). 
The current network queue $Q$ stores two tuples: \pred{+path(@A,C,2,[A,B,C])}
and \pred{+path(@C,A,2,[C,B,A])}, waiting to be delivered to their
destinations (node $A$ and $C$ respectively). 
These two tuples are the result of running \emph{sp2} at node $B$.
We will explain further how configurations are updated based on the updates in
the network queue when introducing the transition rules.
 
\Paragraph{Top-level transitions.}
The small-step operational semantics of a node is denoted $\lstate
\stepsone \lstate', \updates$. From state \lstate, a node takes a step
to a new state $\lstate'$ and generates a set of updates \updates for
other nodes in the network. 
The small-step operational semantics of the entire system is denoted
$\conf \steps{} \conf'$, where $\conf$ and $\conf'$ respectively
  represent the states of all nodes along with the network queue before and
  after the transition.
Figure~\ref{fig:op-semantics} defines the rules
for system state transition. 

\begin{figure}[t!]
\framebox{$\lstate \stepsone \lstate', \updates$}
\begin{mathpar}
\inferrule*[right=Init]{
\updates_\textit{in}=[\ins p_1(@\nodeid, \vec{t}_1), ..., 
\ins p_m(@\nodeid, \vec{t}_m)] \\
 [p_1(@\nodeid,\vec{t}_1), \cdots, p_m(@\nodeid,\vec{t}_m)] =
   \baseof{\prog}
}{
(\nodeid, \emptyset, [], \prog)
\stepsone (\nodeid, \emptyset, \updates_\textit{in},
\prog),  []
}

\and

\inferrule*[right=RuleFire]{
  (\updates_\textit{in}, \updates_\textit{ext}) = 
  \firerules(\nodeid, \tuples, \upd, \Delta\prog)
}{ 
 (\nodeid, \tuples, \upd::\updates, \prog)
\stepsone (\nodeid, \tuples \uplus \upd, \updates\circ\updates_\textit{in},
\prog),  \updates_\textit{ext}
}
\end{mathpar}

\framebox{$\conf \steps{} \conf'$}
\begin{mathpar}
\footnotesize
\inferrule*[right=NodeStep]{\lstate_i \stepsone \lstate'_i, \updates\\
  \forall j\in[1,n] \conjunc j\neq i, \lstate'_j = \lstate_j
 }{
\queue \rhd \lstate_1, \cdots \lstate_n \steps{}
\queue\circ \updates \rhd\lstate'_1, \cdots \lstate'_n
}
\and
\inferrule*[right=DeQueue]{
\queue = \queue'\oplus\queue_1\cdots \oplus \queue_n
\\
  \forall j\in[1,n] \\ \lstate'_j = \lstate_j\circ\queue_j
}{
\queue \rhd \lstate_1, \cdots \lstate_n \steps{}
\queue'\rhd\lstate'_1, \cdots \lstate'_n
}\end{mathpar}

 \framebox{$\firerules(\nodeid, \tuples, \upd, \Delta\prog)= 
(\updates_\textit{in}, \updates_\textit{ext})$
}

 \begin{mathpar}
 \inferrule*[right=Empty]{ }{
 \firerules(\nodeid, \tuples, \upd, []) =
 ([], [])
  }
\and
\inferrule*[right=Seq]{
 \firesingle(\nodeid, \tuples, \upd, \Delta\rules) =
 (\tuples', \updates_\textit{in1}, \updates_\textit{ext1})
 \\
 \firerules(\nodeid, \tuples', \upd, \Delta\prog) = 
 (\updates_\textit{in2}, \updates_\textit{ext2})
  }{
 \firerules(\nodeid, \tuples, \upd,  (\Delta r, \Delta\prog)) = 
 (\updates_\textit{in1}\circ \updates_\textit{in2},
 \updates_\textit{ext1}\circ \updates_\textit{ext2})
 }
 
 \end{mathpar}

\caption{Operational Semantics}
\label{fig:op-semantics}
\end{figure}

\begin{packeditemize}
\item {\bf Global state transition ($\conf \steps{} \conf'$).} 
Rule \rulename{NodeStep} states that the system takes a step when one node
takes a step. As a result, the updates
generated by node $i$ are appended to the end of the network queue. We use
$\circ$ to denote the list append operation. 
Rule \rulename{DeQueue} applies when a node receives
updates from the network. We write $\queue_1\oplus\queue_2$ to denote
a merge of two lists. 

Any node can dequeue updates
sent to it and append those updates to the update list in its
local state. Here, we overload the $\circ$ operator, and write
$\lstate\circ\queue$ to denote a new state, which is the same as
\lstate, except that the update list is the result of appending $\queue$
to the update list in \lstate.

\item {\bf Local state transition ($\lstate \stepsone
\lstate', \updates$).} 
Rule \rulename{Init} applies when the program
starts to run. Here, only base rules---rules that do not have
a rule body---can fire. The auxiliary function \baseof{\prog}
returns all the base rules in \prog. In the resulting state, the
internal update list ($\updates_\textit{in}$) contains all the
insertion updates located at \nodeid, and the
external update list ($\updates_\textit{ext}$) contains only updates 
meant to be stored at a node different from $\nodeid$. In this case,
it is empty.
Rule \rulename{RuleFire} (Figure~\ref{fig:op-semantics}) computes new
updates based on the program and the first update in the update list.
It uses a relation
$\firerules$, which processes an
update \upd, and returns a pair of update lists, one for node \nodeid
itself, the other for other nodes. 
The last argument for $\firerules$, $\Delta\prog$, transforms every
rule $r$ in the program $\prog$ into a \emph{delta} rule, $\Delta r$,
for $r$, which we explain when we discuss incremental maintenance.

After \upd is processed, the
database of \nodeid is updated with the update \upd
($\tuples\uplus\upd$). The $\uplus$ operation increases (decreases)
the reference count of $P$ in \tuples by 1, when \upd is an insertion
(deletion) update $\ins P$ ($\del P$).  The update list in
the resulting state is augmented with the new updates generated from
processing \upd. 

\item {\textbf{Fire rules} ($\firerules(\nodeid, \tuples, \upd, \Delta\prog)= 
(\updates_\textit{in}, \updates_\textit{ext})$).}
Given one update, we fire rules in the program \prog that are affected
by this update. Rule \rulename{Empty} is the base
case where all rules have been fired, so we directly return two empty
sets. Given a program with at least one rule ($\Delta r,
\Delta\prog$), rule \rulename{Seq} first fires the rule $\Delta r$,
then recursively calls itself to process the rest of the rules in
$\Delta\prog$. The resulting updates are the union of the updates from
firing $\Delta r$ and $\Delta\prog$.

\end{packeditemize}

\noindent Given the example scenario in Figure~\ref{fig:scenario}, now node $A$
dequeues the update \pred{+path(@A,C,2,[A,B,C])} from the network queue $Q$ at
the top of Figure~\ref{fig:scenario}, and
puts it into the unprocessed update list $\updates_A$ (rule {\sc DeQueue}). Node
$A$ then locally processes the update by firing all rules that are triggered by
the update, and generates new updates $\updates_{in}$ and $\updates_{ext}$.
In the resulting state, the local
state of node $A$ ($\tuples_A$) is updated with \pred{path(@A,C,2,[A,B,C])}, and
$\updates_A$ now includes $\updates_{in}$. The network queue is also
updated to include $\updates_{ext}$ (rule {\sc NodeStep}).

Our operational semantics does not specify the time gaps between two consecutive
reductions and, therefore, does not determine time points as associated with
a concrete trace---such as $\conf \steps{\tau} \conf'$, where $\tau$ represents the time at
which a concrete transition
takes place. Instead, a trace (without time points) generated by the operational
semantics---e.g., $\conf
\steps{} \conf'$---is an abstraction of all its corresponding annotations with time points that satisfy
monotonicity. In our assertions and proofs, we use time points only to specify a
relative order between events on a specific trace, so their concrete values are
irrelevant.

\Paragraph{Incremental maintenance.}
Now we explain in more detail how the database of a node is maintained
incrementally by processing updates in its internal update list
$\updates_\textit{in}$ one at a time.  
Following the strategy proposed in declarative
networking~\cite{declareNetworks}, the rules in a \Langsec program are rewritten
into \emph{$\Delta$ rules}, which can
efficiently generate all the updates triggered by one update.  For any
given rule \rules that contains $k$ body tuples, $k$ $\Delta$ rules of
the following form are generated, one for each $i\in[1,k]$.\smallskip
\[
\small\Delta p(\halist)\derives
 p^\nu_1(\balist_1), ..., p^\nu_{i-1}(\balist_{i-1}), 
\Delta p_i(\balist_i),
\\~~~~~~~~~~~~~~~~~~~~~~~
p_{i+1}(\balist_{i+1}), ..., p_k(\balist_k),
a_1, ..., a_m
\]
\smallskip

\noindent $\Delta p_i$ in the body denotes the update currently being
 considered.
 $\Delta p$ in the head
 denotes new updates that are generated as the result of firing
 this rule.
 Here $p^\nu_i$ denotes a tuple of name $p_i$ in the database \tuples or the
 internal update list $\updates_{in}$. In comparison, $p_i$ (without $\nu$) denotes a tuple of name
 $p_i$ only in \tuples.
For example, the $\Delta$ rules for {\em
  sp2} are:
\[
\hspace{-15pt}\small
\begin{array}{l@{~~}l}
\textit{sp2a}&\pred{$\Delta$path}(@z, d, c, p) \derives \pred{$\Delta$link}(@s, z, c1), \pred{path}(@s, d, c2, p1), 
c := c1 + c2, p := z{::}p1. 
\\
\textit{sp2b}&\pred{$\Delta$path}(@z, d, c, p) \derives \pred{link$^\nu$}(@s, z, c1), \pred{$\Delta$path}(@s, d, c2, p1), 
c := c1 + c2, p := z{::}p1. 
\end{array}
\]

\noindent Rules {\em sp2a} and {\em sp2b} are $\Delta$ rules triggered
by updates of the \pred{link} and \pred{path} relation
respectively. For instance, when node $A$ processes
$\pred{+path(@A,C,2,[A,B,C])}$, only rule {\em sp2b} is fired.
In this step, $\pred{path}^\nu$ includes the tuple 
 $\pred{path(@A,C,2,[A,B,C])}$, while $\pred{path}$ does not.
On the other hand, $\pred{link}^\nu$ and $\pred{link}$ denote the same set of
tuples, because $\updates_{in}$ does not contain any tuple of name $\pred{link}$.
The rule evaluation then generates $\pred{+path(@B,C,3,[B,A,B,C])}$, which will 
be communicated to node $B$ and further triggers rule {\em sp2b} at node $B$. Such update propagates
until no further new tuples are generated. 

\Paragraph{Rule Firing.} 
We present in Figure~\ref{fig:delta:rules:single} the set of
rules for firing a single $\Delta$ rule given an insertion
update. 
We write $\tuples^\nu$ to denote the table resulted from
updating \tuples with the current update: $\tuples^\nu
=\tuples\uplus\upd$. 
\begin{figure*}[t!]
\framebox{$\firesingle(\nodeid, \tuples,  \upd , \Delta\rules)
  = (\tuples', \updates_\textit{in}, \updates_\textit{ext})$}
\begin{center}
 \[
 \inferrule*[right=InsExists]{
 (n, q_i(\vec{t})) \in\tuples }{
   \firesingle(\nodeid, \tuples,  \ins q_i(\vec{t}) , \Delta\rules)
   = (\tuples, [], [])
 }
 \]

 \figspace

\[
\inferrule*[right=InsNew]{
\Delta\rules = \Delta p(@\nodeid_1, \alist)\derives 
  \cdots,\Delta q_i(\balist_i) \cdots
  \\
q_i(\vec{t}) \notin\tuples 
\\
 \alist~\mbox{does not contain any aggregate}
\\
\Sigma = \substof(\tuples^\nu, \tuples, \rules, i, \vec{t})
\\
 \Sigma' = \select(\Sigma, \tuples^\nu)
\\
 \updates = \genupd(\Sigma, \Sigma', p, \tuples^\nu)
\\ 
\mbox{if}~\nodeid_1 = \nodeid~\mbox{then}~\updates_i = \updates,
\updates_e =[]
~\mbox{otherwise}~ \updates_i = [], \updates_e = \updates
 }{
  \firesingle(\nodeid, \tuples,  \ins q_i(\vec{t}) , \Delta\rules)
  = (\tuples, \updates_i, \updates_e)
}
\]

\figspace

 \[
 \inferrule*[right=InsAggSame]{
 \Delta\rules = \Delta p(@\nodeid, \alist)\derives 
   \cdots,\Delta q_i(\balist_i) \cdots
   \\
 q_i(\vec{t}) \notin\tuples 
 \\
  \alist~\mbox{contains an aggregate \agf}
 \\
  \{\sigma_1,\cdots, \sigma_k\} = \substof(\tuples^\nu, \tuples,
  \rules, i, \vec{t})
 \\
 \tuples' = \tuples \uplus \{p_\textit{agg}(@\nodeid, \sigma_1(\alist)),
 \cdots, p_\textit{agg}(@\nodeid, \sigma_k(\alist))\}
 \\
 \aggof(p,\agf ,\tuples') = p(@\nodeid,\vec{s})
 \\ 
   p(@\nodeid,\vec{s}) \in \tuples
  }{
   \firesingle(\nodeid, \tuples,  \ins q_i(\vec{t}) , \Delta\rules)
   = (\tuples', [], [])
 }
 \]

\figspace

 \[
 \inferrule*[right=InsAggUpd]{
 \Delta\rules = \Delta p(@\nodeid, \alist)\derives 
   \cdots,\Delta q_i(\balist_i) \cdots
   \\
 q_i(\vec{t}) \notin\tuples 
 \\
  \alist~\mbox{contains an aggregate \agf}
 \\
  \{\sigma_1,\cdots, \sigma_k\} = \substof(\tuples^\nu, \tuples,
  \rules, i, \vec{t})
 \\
 \tuples' = \tuples \uplus \{p_\textit{agg}(@\nodeid, \sigma_1(\alist)),
 \cdots, p_\textit{agg}(@\nodeid, \sigma_k(\alist))\}
 \\
 \aggof(p,\agf ,\tuples') = p(@\nodeid,\vec{s})
 \\ 
   p(@\nodeid,\vec{s}_1) \in \tuples
 \\\mbox{$\vec{s}$ and $\vec{s}_1$ share the same key but different
   aggregate value}
  }{
   \firesingle(\nodeid, \tuples,  \ins q_i(\vec{t}) , \Delta\rules)
   = (\tuples', [\del p(@\nodeid, \vec{s}_1), \ins p(@\nodeid, \vec{s})], [])
 }
 \]

 \figspace
\[
\inferrule*[right=InsAggNew]{
\Delta\rules = \Delta p(@\nodeid, \alist)\derives 
  \cdots,\Delta q_i(\balist_i) \cdots
\\
q_i(\vec{t}) \notin\tuples 
\\
 \alist~\mbox{contains an aggregate \agf}
\\
 \{\sigma_1,\cdots, \sigma_k\} = \substof(\tuples^\nu,  \tuples,
 \rules, \vec{t})
\\
\tuples' = \tuples \uplus \{p_\textit{agg}(@\nodeid, \sigma_1(\alist)),
\cdots, p_\textit{agg}(@\nodeid, \sigma_k(\alist))\}
\\
\aggof(p,\agf ,\tuples') = p(@\nodeid,\vec{s})
\\ 
 \nexists p(@\nodeid,\vec{s}') \in \tuples
\\\mbox{such that $\vec{s}$ and $\vec{s'}$ share the same key but different
  aggregate value}
 }{
  \firesingle(\nodeid, \tuples,  \ins q_i(\vec{t}) , \Delta\rules)
  = (\tuples', [\ins p(@\nodeid, \vec{s})], [])
}
\]
\end{center}
\caption{Insertion rules for evaluating a single $\Delta$ rule}
\label{fig:delta:rules:single}
\end{figure*}
Rule \rulename{InsExists} specifies the case where the tuple to be inserted
(i.e. $q_i(\vec{t})$) already exists. We do not need to further propagate the
update. 
Rule \rulename{InsNew} handles the case where new updates are
generated by firing rule \rules. In order to fire a rule \rules, we need to map
its bodies to concrete tuples in the database or the update list. We use an auxiliary function
$\substof(\tuples^\nu, \tuples, \rules, i, \vec{t})$ to extract the complete list
of substitutions for variables in the rule.
Here $i$ and $\vec{t}$
indicate that $q_i(\vec{t})$ is the current update, where $q_i$ is the
$i^{th}$ body tuple of rule $\rules$. Every
substitution $\sigma$ in that set is a
general unifier of the body tuples and constraints. Formally: 
\begin{tabbing}
(1) $\vec{t} = \sigma(\balist_i)$, \quad \= \\
(2) $\forall j\in[1,i-1], \exists \vec{s}, 
  \vec{s} = \sigma(\balist_j)$ and
 $q_j(\vec{s}) \in \tuples^\nu$ \\
 (3) $ \forall j\in[i+1,n], \exists \vec{s}, 
  \vec{s} = \sigma(\balist_j)$ and 
 $q_j(\vec{s}) \in \tuples$ \quad \\
(4) $ \forall k\in[1, m], \sigma[a_k]$ is true
\end{tabbing}
We write $[a]$ to denote the constraint that $a$ represents.
When $a$ is an assignment (i.e., $x:=f(\vec{t})$), $[a]$ 
is the equality constraint $x=f(\vec{t})$; otherwise, $[a]$ is $a$.

When multiple tuples with the same key are derived using a
rule, a selection function \select is introduced to decide which substitution to
propagate. In \Langsec run time, similar to a relational database, a key value of a
stored tuple $p(\vec{t})$ uniquely identifies that tuple. When a different tuple
$p(\vec{t'})$ with the same key is derived, the old value $p(\vec{t})$ and any
tuple derived using it need to be deleted.  For instance,
we can demand that each pair of nodes in the network have a unique path between
them. This is equivalent to designating the first two arguments of $\pred{path}$
as its key.
As a result, $\pred{path(A,B,1,[A,B])}$ and $\pred{path(A,B,2,[A,D,B])}$ cannot
both exist in the database.  

We also use a \genupd function to generate appropriate updates based on the
selected substitutions. It may generate deletion updates in addition
to an insertion update of the new value. For example, assume that
$\pred{path(A,B,3,[A,C,D,B])}$ is in $\tuples^\nu$.  If we were to
choose $\pred{path(A,B,1,[A,B])}$ because it appears earlier in the
update list, then \genupd returns $\{\ins\pred{path(A,B,1,[A,B])},
\del\pred{path(A,B,3,[A,C,D,B])}\}$.  We leave the definitions of
\select and \genupd abstract here, as there are many possible
strategies for implementing these two functions. Aside from the
strategy of picking the first update in the queue (illustrated above),
another possible strategy is to pick the last, as it is the freshest.
Once the strategy of \select is fixed, \genupd is also fixed.
However, the only relevant part to the logic we introduce later is
that the substitutions used for an insertion update come from the
$\substof$ function, and that the substitutions satisfy the property
we defined above. In other words, our program logic can be applied to
a number of different implementation of \select and \genupd.

The rest of the rules in Figure~\ref{fig:delta:rules:single} deal with
generating an aggregate tuple. Rule {\sc InsAggNew} applies when the
aggregate is generated for the first time. We only need to insert the new
aggregate value to the table. Additional rules (i.e. {\sc InsAggSame} and {\sc
  InsAggUpd}) are required to handle aggregates
where the new aggregate is the same as the old one or replaces the old one.
To efficiently implement aggregates, for each tuple $p$ that has an aggregate
function in its arguments, there is an internal tuple $p_\textit{agg}$ that
records all candidate values of $p$. When there is a change to the candidate
set, the aggregate is re-computed.  For example, $\pred{bestpath}_\textit{agg}$
maintains all candidate \pred{path} tuples.

We also require that the location specifier of a rule head containing an
aggregate function be the same as that of the rule body. With this
restriction, the state of an aggregate is maintained in one single
node. If the result of the aggregate is needed by a remote node, we
can write an additional rule to send the result after the aggregate
is computed. 

\begin{figure}[t!]
\begin{mathpar}
\inferrule*[right=DelExists]{
(n, q_i(\vec{t})) \in\tuples \\ n > 1}{
  \firesingle(\nodeid, \tuples,  \del q_i(\vec{t}) , \Delta\rules)
  = (\tuples, [], [])
}\and
\inferrule*[right=DelNew]{
\Delta\rules = \Delta p(@\nodeid_1, \alist)\derives 
  \cdots,\Delta q_i(\balist_i) \cdots
  \\
  (1, q_i(\vec{t})) \in\tuples 
\\
 \alist~\mbox{does not contain any aggregate}
\\
 \{\sigma_1, \cdots, \sigma_k\} = \select(\substof(\tuples^\nu, \tuples,
 \rules, i, \vec{t}), \tuples^\nu)
\\
 \updates = [\del p(@\nodeid_1, \sigma_1(\alist)), \cdots, \del
   p(@\nodeid_1, \sigma_k(\alist))]
\\
\\ \mbox{if}~\nodeid_1 = \nodeid~\mbox{then}~\updates_i = \updates,
\updates_e =[]
~\mbox{otherwise}~ \updates_i = [], \updates_e = \updates
 }{
  \firesingle(\nodeid, \tuples,  \del q_i(\vec{t}) , \Delta\rules)
  = (\tuples, \updates_i, \updates_e)
}\and
\inferrule*[right=DelAggSame]{
\Delta\rules = \Delta p(@\nodeid, \alist)\derives 
  \cdots,\Delta q_i(\balist_i) \cdots
  \\
(1, q_i(\vec{t})) \in\tuples 
\\
 \alist~\mbox{contains an aggregate \agf}
\\
 \{\sigma_1,\cdots, \sigma_k\} = \substof(\tuples^\nu, \tuples,
 \rules, i, \vec{t})
\\
\tuples' = \tuples \backslash \{p_\textit{agg}(@\nodeid, \sigma_1(\alist)),
\cdots, p_\textit{agg}(@\nodeid, \sigma_k(\alist))\}
\\
\aggof(p,\agf ,\tuples') = p(@\nodeid,\vec{s})
\\ 
  p(@\nodeid,\vec{s}) \in \tuples
 }{
  \firesingle(\nodeid, \tuples,  \del q_i(\vec{t}) , \Delta\rules)
  = (\tuples', [], [])
}\and
\inferrule*[right=DelAggUpd]{
\Delta\rules = \Delta p(@\nodeid, \alist)\derives 
  \cdots,\Delta q_i(\balist_i) \cdots
  \\
(1, q_i(\vec{t})) \in\tuples 
\\
 \alist~\mbox{contains an aggregate \agf}
\\
 \{\sigma_1,\cdots, \sigma_k\} = \substof(\tuples^\nu, \tuples,
 \rules, i, \vec{t})
\\
\tuples' = \tuples \backslash \{p_\textit{agg}(@\nodeid, \sigma_1(\alist)),
\cdots, p_\textit{agg}(@\nodeid, \sigma_k(\alist))\}
\\
\aggof(p,\agf ,\tuples') = p(@\nodeid,\vec{s})
\\ 
  p(@\nodeid,\vec{s}_1) \in \tuples
\\\mbox{$\vec{s}$ and $\vec{s}_1$ share the same key but different
  aggregate value}
 }{
  \firesingle(\nodeid, \tuples,  \del q_i(\vec{t}) , \Delta\rules)
  = (\tuples', [\del p(@\nodeid, \vec{s}_1), \ins p(@\nodeid, \vec{s})], [])
}\and
\inferrule*[right=DelAggNone]{
\Delta\rules = \Delta p(@\nodeid, \alist)\derives 
  \cdots,\Delta q_i(\balist_i) \cdots
  \\
(1, q_i(\vec{t})) \in\tuples 
\\
 \alist~\mbox{contains an aggregate \agf}
 \\
 \{\sigma_1,\cdots, \sigma_k\} = \substof(\tuples^\nu, \tuples,
 \rules, i, \vec{t})
\\
\tuples' = \tuples \backslash \{p_\textit{agg}(@\nodeid, \sigma_1(\alist)),
\cdots, p_\textit{agg}(@\nodeid, \sigma_k(\alist))\}
\\
\aggof(p,\agf ,\tuples') = \sf{NULL}
 }{
  \firesingle(\nodeid, \tuples,  \del q_i(\vec{t}) , \Delta\rules)
  = (\tuples', [\del p(@\nodeid, \vec{s'})], [])
}
\end{mathpar}
\caption{Deletion rules for evaluating a single $\Delta$ rule}
\label{fig:delta-rules-single-del}
\end{figure}

 Rule \rulename{InsAggSame} applies when the new aggregates is
 the same as the old one. In this case, only the candidate set is
 updated, and no new update is propagated. Rule \rulename{InsAggUpd}
 applies when there is a new aggregate value. In this case, we need to
 generate a deletion update of the old tuple before inserting the new
 one.

Figure~\ref{fig:delta-rules-single-del} summaries the deletion rules.
When the tuple to
be deleted has multiple copies, we only reduce its reference count. 
The rest of the rules 
are the dual of the corresponding insertion rules. 
We revisit the example in Figure~\ref{fig:scenario} to illustrate
how incremental maintenance is performed on the shortest-path program.  Upon
receiving \pred{+path(@A,C,2,[A,B,C])}, $\Delta$ rule {\em sp2b} will be
triggered and generate a new update \pred{+path(@B,C,3,[B,A,B,C])},
which will be included in $\updates_{ext}$ as it is destined to a
remote node $B$ (rule {\sc InsNew}). The $\Delta$ rule for {\em sp3}
will also be triggered, and generate a new update
\pred{+bestPath(@A,C,2,[A,B,C])}, which will be included in
$\updates_{in}$ (rule {\sc InsAggNew}).  After evaluating the $\Delta$
rules triggered by the update \pred{+path(@A,C,2,[A,B,C])}, we have
$\updates_{in} = \{\pred{+bestPath(@A,C,2,[A,B,C])}\}$ and
$\updates_{ext} = \{\pred{+path(@B,C,3,[B,A,B,C])}\}$.    In addition,
$\pred{bestpath}_\textit{agg}$, the auxiliary relation that maintains
all candidate tuples for \pred{bestpath}, is also updated to reflect
that a new candidate tuple has been generated. It now includes
$\pred{bestpath}$\pred{(@A,C,2,[A,B,C])}.  

\Paragraph{Discussion.} 
The semantics introduced here will
not terminate for programs with a cyclic derivation of the same
tuple, even though set-based semantics will. Most routing
protocols do not have such issue (e.g., cycle detection is well-adopted in routing protocols). Our prior work~\cite{ppdp11} has
proposed improvements to solve this issue. It is a straightforward
extension to the current semantics and is not crucial for
demonstrating the soundness of the program logic we
develop. 

The operational semantics is correct if the results are the same as
one where all rules reside in one node and a global fixed point is
computed at each round. The proof of correctness is out of the scope
of this paper. 
We are working on
correctness definitions and proofs for variants of PSN algorithms. Our
initial results for a simpler language
can be found in~\cite{ppdp11}. \Langsec additionally allows
aggregates, which are not included in~\cite{ppdp11}. The soundness of
our logic only depends on the specific evaluation strategy implemented
by the compiler, and is orthogonal to the correctness of the
operational semantics. Updates to the operational semantics is likely
to come in some form of additional bookkeeping in the
representation of tuples, which we believe will not affect the overall
structure of the program logic; as these metadata are irrelevant to the
logic. \reviewer{The last
  sentence...is a bit mysterious though. Is this a good thing? A bad
  thing? Is it just unavoidable?}

\section{A Program Logic for \Langsec}
\label{sec:verification}
To verify correctness of secure routing protocols encoded in \Langsec, we
introduce a program logic for \Langsec.
The program logic enables us to prove program
  invariants---that is, properties holding throughout the execution of
  \Langsec programs---even if the nodes running the program interact
  with potential attackers, whose behaviors are unpredictable. The
  properties of secure routing protocols that we are interested in are
  all safety properties and can be verified by analyzing programs'
  invariant properties.

\Paragraph{Attacker model.}
We assume \emph{connectivity-bound} network attackers, a variant of
the Dolev-Yao network attacker model. The attacker can perform
cryptographic operations with correct keys, such as
encryption, decryption, and signature generation, but is not allowed to eavesdrop or
intercept packets. 
This attacker model manifests itself in our formal system in two places:
(1) the network is modeled as connected nodes, some of which run the \Langsec program
that encodes the prescribed protocol and others are malicious and run arbitrary
\Langsec programs; (2) safety of cryptography is admitted as axioms in our proofs.

\Paragraph{Syntax.}
We use first-order logic formulas, denoted $\varphi$, as property specifications.
The atoms, denoted $A$, include predicates and
term inequalities. 
The syntax of the logic formulas is shown below.
\[\hspace{-15pt}\small
\begin{array}{l@{~~}lcl}
\textit{Atoms}& A &  \bnfdef &
P(\vec{t}) @ (\nodeid, \tau) 
\bnfalt  \pred{send}(\nodeid, \pred{tp}(P, \nodeid', \vec{t})) @
\tau  
\bnfalt   \pred{recv}(\nodeid, \pred{tp}(P,\vec{t})) @ \tau
\\ & & \bnfalt  & 
\pred{honest}(\nodeid, \prog, \tau) 
\bnfalt  t_1 \bop t_2
 \\
 \textit{Formulas} & \varphi& \bnfdef & 
  \top\bnfalt \bot\bnfalt A \bnfalt \varphi_1\conjunc\varphi_2
  \bnfalt \varphi_1\disj\varphi_2
 \bnfalt \varphi_1\imp\varphi_2\bnfalt \neg \varphi
 \bnfalt \forall x. \varphi\bnfalt \exists x.\varphi
 \\
 \textit{Variable Ctx} & \Sigma & \bnfdef & \cdot \bnfalt \Sigma, x
 \qquad\qquad\qquad
 \textit{Logical Ctx} \quad \Gamma ~ \bnfdef ~ \cdot\bnfalt \Gamma, \varphi
\end{array} 
\]

\begin{figure}[t!]
\begin{mathpar}
\inferrule*[right=Cut]{
\Sigma;\Gamma  \vdash \varphi \\
\Sigma;\Gamma, \varphi \vdash \varphi' \\
}{
\Sigma;\Gamma  \vdash \varphi' \\
}\and
\inferrule*[right=Init]{
\varphi \in \Gamma
}{
 \Sigma; \Gamma  \vdash \varphi 
}\and
\inferrule*[right=$\neg$I]{ 
  \Sigma; \Gamma,\varphi \vdash \cdot
}{  \Sigma; \Gamma  \vdash
  \neg\varphi}\and
\inferrule*[right=$\neg$E]{ 
  \Sigma; \Gamma  \vdash \neg\varphi
}{  \Sigma; \Gamma , \varphi \vdash \cdot
  }\and

 \inferrule*[right=$\conjunc$I]{
  \Sigma; \Gamma  \vdash \varphi_1 
\\   \Sigma; \Gamma  \vdash \varphi_2 
 }{ \Sigma; \Gamma  \vdash \varphi_1\conjunc\varphi_2 }\and
 \inferrule*[right=$\conjunc$E]{
 i\in[1,2], 
  \Sigma; \Gamma  \vdash \varphi_1\conjunc\varphi_2  
 }{ \Sigma; \Gamma  \vdash \varphi_i }\and
 \inferrule*[right=$\disj$I]{
 i\in[1,2], 
  \Sigma; \Gamma  \vdash \varphi_i 
 }{ \Sigma; \Gamma  \vdash \varphi_1\disj\varphi_2 }\and
 \inferrule*[right=$\disj$E]{
  \Sigma; \Gamma  \vdash \varphi_1\disj\varphi_2 
 \\
  \Sigma; \Gamma , \varphi_1 \vdash \varphi
 \\  \Sigma; \Gamma , \varphi_2 \vdash \varphi
 }{ \Sigma; \Gamma \vdash \varphi}\and
\end{mathpar}
\begin{mathpar}
\inferrule*[right=$\forall$I]{
 \Sigma, x; \Gamma  \vdash \varphi 
}{ \Sigma; \Gamma  \vdash \forall
  x. \varphi }\and
\inferrule*[right=$\forall$E]{
 \Sigma; \Gamma \vdash \forall
  x. \varphi  \\
}{ \Sigma; \Gamma  \vdash \varphi[t/x]}\and
\inferrule*[right=$\exists$I]{
 \Sigma; \Gamma \vdash \varphi[t/x]
}{ \Sigma; \Gamma  \vdash \exists
  x. \varphi  }\and
\inferrule*[right=$\exists$E]{
 \Sigma; \Gamma  \vdash \exists x.\varphi 
\\
 \Sigma, a; \Gamma , \varphi[a/x] 
 \vdash \varphi'
\\ a~\mbox{is fresh}
}{ \Sigma; \Gamma  \vdash \varphi'}
\end{mathpar}
\vspace{-10pt}
\label{fig:logic-rules}
\caption{Rules in first-order logic.}
\end{figure}

\noindent Predicate $P(\vec{t}) @ (\nodeid,
\tau)$ means that tuple $P(\vec{t})$ is derived at time $\tau$ by
node \nodeid. The first element in $\vec{t}$ is a location identifier
$\nodeid'$, which may be different from $\nodeid$.  When a tuple
$P(\nodeid', ...)$ is derived at node $\nodeid$, it is sent to
$\nodeid'$. This {\em send} action is captured by predicate
$\pred{send}(\nodeid, \pred{tp}(P, \nodeid', \vec{t})) @ \tau$. 
Correspondingly, predicate $\pred{recv}(\nodeid, \pred{tp}(P,\vec{t}))
@ \tau$ denotes that node \nodeid has received a tuple $P(\vec{t})$ at
time $\tau$. A user could determine \pred{send} and \pred{recv} tuples by
inspecting rules whose head tuple locates differently from body tuples. For
example, the head tuple $\pred{path}(@z,d,c,p)$ in the rule \textit{sp2} of the
shortest-path program (Figure~\ref{fig:bestpath}) corresponds to a tuple
$\pred{send}(s,\pred{tp}(path, z, (z,d,c,p)))@t$ in our logic. $\pred{honest}(\nodeid,
\prog(\nodeid), \tau)$ means that
node $\nodeid$ starts to run program $\prog(\nodeid)$ at time $\tau$.
Since predicates take time points as an
argument, we are effectively encoding linear temporal logic (LTL) in
first-order logic~\cite{phd-kamp}. The domain of the time points is
the set of natural numbers. Each time point represents the number of
clock ticks from the initialization of the system.

\Paragraph{Logical judgments.}
The logical judgments in our program logic use two contexts: context $\Sigma$,
which contains all the free variables; and context $\Gamma$, which contains logical assumptions.\medskip

(1) $\Sigma;\Gamma\vdash \varphi$
\qquad\qquad
(2) $\Sigma; \Gamma \vdash \runby{\prog}{i} : \{i, y_b, y_e\}.\varphi(i, y_b,y_e)$\medskip

\reviewer{I guess you're already
talking about (2), not still about (1)}
\noindent Judgment (1) states that $\varphi$ is provable given the assumptions
in $\Gamma$.  Judgment (2) is an assertion about
\Langsec programs, i.e., a program invariant. We write $\varphi(\vec{x})$
when $\vec{x}$ are free in $\varphi$.
$\varphi(\vec{t})$ denotes the resulting formula of substituting $\vec{t}$ for
$\vec{x}$ in  $\varphi(\vec{x})$. Recall that
 \prog is parametrized over the
identifier of the node it runs on. 

The program invariant is parametrized over not only the node ID $i$, but also the
starting point of executing the program ($y_b$) and a later time point $y_e$.
Judgment (2) states that any trace \trace containing
the execution of a program \prog by a node \nodeid, starting
at time $\tau_b$, satisfies $\varphi(\nodeid, \tau_b, \tau_e)$, for any
time point $\tau_e$ later than $\tau_b$. 
Note that the trace could also contain threads that run malicious programs.
Since $\tau_e$ is any time after $\tau_b$ (the time \prog starts), 
$\varphi$ is an invariant property of \prog. 

\begin{figure}[t!]
\framebox{$\footnotesize\Sigma; \Gamma \vdash \runby{\prog}{i} : \{i, y_b, y_e\}.\varphi(i,
  y_b,y_e)$}
\begin{mathpar}
 \mprset{flushleft}
 \inferrule*[right=Inv]{
 \forall r\in\ruleof{\prog}, ~~~~~~~~
   (r = h(\vec{v}) \derives p_1(\vec{s}_1),..., p_m(\vec{s}_m), 
   q_1(\vec{u}_1), ..., q_n(\vec{u}_n), a_1,..., a_k)
 \\
 \quad \Sigma;\Gamma\vdash 
  \forall i, \forall t, \forall\vec{y},
  ~~~~~~~~~~ (~\vec{y} = \fv(r))
\\\\
\\  \\
\bigwedge_{j\in[1,m]} (p_j(\vec{s}_j)@(i, t) {\conjunc} \varphi_{p_j}(i,
  t, \vec{s}_j)) {\conjunc}
\\\\
\\  \\
 \bigwedge_{j\in[1,n]} \pred{recv}(i, \pred{tp}(q_j, \vec{u}_j))@ t {\conjunc} 
\\~~~~~~~~~~\mathlarger{\mathlarger{\mathlarger{\mathlarger{\imp}}}} ~~~~~~~~~~
\varphi_h(i, t, \vec{v})
\\\\
\\  \\
\bigwedge_{j\in[1,k]} [a_j]
\\\\\\
 \forall p\in\headof{\prog},
   ~\varphi_p~\mbox{is closed under trace extension} \\
 }{
 \Sigma; \Gamma \vdash \runby{\prog}{i} : 
 \{i, y_b, y_e\}.\bigwedge_{p\in\headof{\prog}}\forall t,
\forall\vec{x}, y_b\leq t<y_e\conjunc
  p(\vec{x})@(i, t)
   \imp \varphi_{p}(i, t, \vec{x})
 }
\end{mathpar}
\framebox{\small$\Sigma;\Gamma\vdash \varphi$} 

\[
\inferrule*[right=Honest]{
 \Sigma; \Gamma \vdash \runby{\prog}{i} : \{i, y_b, y_e\}.\varphi(i, y_b,y_e)
~~ \Sigma; \Gamma \vdash \pred{honest}(\nodeid, \prog(\nodeid), t)
}{
\Sigma;\Gamma \vdash \forall t', t'>t, \varphi(\nodeid, t, t')
}\]
\caption{Rules in program logic}
\label{fig:program:logic}
\end{figure}

\Paragraph{Inference rules.}
The inference rules of our program logic include all standard
first-order logic ones (e.g. Modus ponens), shown in
Figure~\ref{fig:logic-rules}.  Reasoning about the
ordering between time points are carried out in first-order logic
using theory on natural numbers (in Coq, we use Omega). We choose
first-order logic because it is better supported by proof assistants
(e.g. Coq).

In addition, we introduce two key rules (Figure~\ref{fig:program:logic}) 
into our proof system. 
Rule \rulename{Inv} proves an invariant property of a program
\prog. 
The program invariant takes on a specific form as the conjunction of all the
invariants of the tuples derived by \prog, and means that if any head tuple is derived by
\prog, then its associated property should hold; formally:
$\forall t,\forall\vec{x}, y_b\leq t<y_e\conjunc p(\vec{x})@(i, t) \imp
\varphi_{p}(i, t, \vec{x})$, where $p$ is the name of the head tuple, and $\varphi_p(i, t, \vec{x})$ is an invariant property
associated with $p(\vec{x})$. For example, $p$ can be \pred{path}, and
$\varphi_p(i, t, \vec{x})$ be that every link in argument \textit{path}
must have existed in the past. In the \rulename{INV} rule, the function $\mathit{rlOf(\prog)}$ returns rules
generating derivation tuples for a given program, and the function $\mathit{\fv(r)}$
returns all free variables in a given rule.

Intuitively, the premises of \rulename{Inv} need to establish that
\textit{each} derivation rule's body tuples  and its associative invariants together
imply the invariant of the rule's head tuple. 
For each derivation rule $r$ in
\prog, we assume that the body of $r$ is arranged so that the first
$m$ tuples (i.e. $p_1(\vec{s}_1),..., p_m(\vec{s}_m)$) are derived by \prog ,
the next $n$ tuples (i.e. $q_1(\vec{u}_1), ..., q_n(\vec{u}_n)$) are received from
the network, and constraints (i.e. $a_1,..., a_k$) constitute the rest of the body. 
For tuples derived by \prog (i.e. $p_j$'s), we can assume
that their invariants $\varphi_{p_j}$ hold at time $t$. On the other hand,
properties of received tuples (i.e. $q_j$) are excluded from the premises,
as in an adversarial environment, messages from the network are not trusted
by default. 

Each premise of the \rulename{INV} rule provides the strongest
  assumption that allows us to prove the conclusion in that premise. In
  most cases, arithmetic constraints are enough for proving the
  invariant. But in some special cases---for example, the invariant explicitly
  specifies the existence of a received tuple---the
  predicate representing the action of a tuple receipt is needed in
  the assumption. In other words, $\varphi_{p_j}$ is the inductive
  hypothesis in this inductive proof. In our case study, we frequently
  need to invoke the inductive hypothesis to complete the proof.

We make sure that each tuple in an \Langsec program is either derived
locally or received from the network, but not both. For a program that
violates this property, the user can rewrite the program by creating a copy tuple of a
different name for the tuple that can be both
derived locally or received from the network. For example, the \pred{path} tuple in the shortest-path program in
Figure~\ref{fig:bestpath} could be both derived locally (rule \textit{sp1}) and
received from a remote node (rule \textit{sp2}). The user could rewrite the head
tuple \pred{path} in \textit{sp2} to \pred{recvPath} to differentiate it from
\pred{path}. In this way, the invariant property associated with the \pred{path}
tuple can be trusted and used in the proof of the program invariant.

We also require that an invariant $\varphi_p$ be
closed under trace extension. Formally: 
if
$\trace\vDash
\varphi(\nodeid,t,\vec{s})$ and $\trace$ is a prefix of $\trace'$, then
$\trace'\vDash \varphi(\nodeid,t,\vec{s})$. For instance, the property
that node \nodeid has received a tuple $p$ before time $t$ is closed
under trace extension, while the property that node \nodeid never sends $p$
to the network is not closed under trace extension.  
We do not allow invariants to be specified over base tuples. The
  \rulename{INV} rule cannot be used to derive properties of base
  rules (e.g., $\pred{link}$), because the function $\mathit{rlOf()}$ only
  returns rules for derivation tuples.

As an example, we use \rulename{INV} to prove a simple program
invariant of the shortest-path program in
Figure~\ref{fig:bestpath}. The property is specified as

\parbox{0.9\textwidth}{%
  \noindent $\varphi_\textit{sp}$ = $\runby{\prog}{x}: \{x, y_b, y_e\}. \\
~~~~~~~~~~~~~~~~~~~~~  (\forall t, \forall y, \forall c, \forall pt,
  y_b\leq t<y_e\conjunc \\
~~~~~~~~~~~~~~~~~~~~~~~~~~~~  \pred{path}(x,y,c,pt)@(x, t) \imp \\
~~~~~~~~~~~~~~~~~~~~~~~~~~~~~~~~ (\exists z,c',
\pred{link}(x,z,c')@(x,t) \disj \\
~~~~~~~~~~~~~~~~~~~~~~~~~~~~~~~~~~~~~~~~~ \pred{link}(z,x,c'@(z,t))) \conjunc \\
~~~~~~~~~~~~~~~~~~~~~(\forall t, \forall y, \forall c, \forall pt,
  y_b\leq t<y_e\conjunc \\
~~~~~~~~~~~~~~~~~~~~~~~~~~~~  \pred{bestPath}(x,y,c,pt)@(x, t) \imp \ttrue
$ 
}\medskip

\noindent Intuitively, $\varphi_\textit{sp}$ specifies an invariant
property for the \pred{path} tuple, which says a \pred{path} tuple must imply a
\pred{path} tuple to/from the direct neighbor. $\varphi_\textit{sp}$ also assigns $\ttrue$ as the invariant property for \pred{bestPath} tuples. The proof
is established using \rulename{INV} (Figure~\ref{fig:proof}). The whole proof has
three premises, each corresponding to a rule in the shortest-path
program in Figure~\ref{fig:bestpath}. For example, in the second
premise corresponding to \textit{sp2}, we include the local \pred{link}
tuple and the received \pred{path} as well as constraints in the assumption,
while leaving out the invariant property of the
\pred{path} tuple, because a received \pred{path} tuple should not be trusted in
an adversarial environment.

\begin{figure}[t!]
\begin{mathpar}
 \mprset{flushleft}
 \inferrule*[right=Inv]{
\\\\
\\  \\
\\\\
 \Sigma;\Gamma\vdash 
  \forall s, \forall d, \forall c, \forall t,
\\\\
~~~~~~~~~~(\pred{link}(s,d,c)@(s,t) \conjunc p = [s,d]) \imp
\\\\
~~~~~~~~~~~~~~(\exists z, c', \pred{link}(s,z,c')@(s,t) \disj \pred{link}(z,s,c')@(z,t))
\\\\
\\  \\
\\\\
\Sigma;\Gamma\vdash 
\forall s, \forall d, \forall c1, \forall c2, \forall p1, \forall z, \forall t,
\\\\
~~~~~~~~~~(\pred{link}(s, z, c1)@(s,t) \conjunc \pred{recv}(s, \pred{tp}(\pred{path},s,d,c2,p1))@t \conjunc
\\\\
~~~~~~~~~~c = c1 + c2 \conjunc p = z{::}p1) \imp
\\\\
~~~~~~~~~~~~~~(\exists z'',c'', \pred{link}(z,z'',c'')@(z,t) \disj
\pred{link}(z'',z,c')@(z'',t))
\\\\
\\  \\
\\\\
 \Sigma;\Gamma\vdash \ttrue
\\\\
 }{
 \Sigma; \Gamma \vdash \varphi_\textit{sp}
 }
\end{mathpar}
\caption{Proof of $\varphi_\textit{sp}$}
\label{fig:proof}
\end{figure}

The \rulename{Honest} rule proves properties of the entire system
based on the program invariant.  If $\varphi(i, y_b, y_e)$
is the invariant of \prog, and a node \nodeid runs the program \prog
at time $t_b$, then any trace \emph{containing} the execution of this
program satisfies $\varphi(\nodeid, t_b, t_e)$, where $t_e$ is a time
point after $t_b$.  \Langsec programs never terminate: after the last
instruction, the program enters a stuck state. The \rulename{Honest}
rule is applied to honest principles (nodes) that execute the
prescribed protocols. The invariant property of an honest node holds
even when it interacts with other malicious nodes in the network,
which is required by the soundness of the inference rules.  We explain
in more detail next.

\Paragraph{Soundness.}
We prove the soundness of our logic with regard to
the trace semantics. First, we define the
trace-based semantics for our logic and judgments in
Figure~\ref{fig:semantics}. Different from semantics of first-order logic, in
our semantics,
formulas are interpreted on a trace
$\trace$. We elide the rules for first-order logic
connectives. A tuple $ P(\vec{t})$ is derivable by node \nodeid at
time $\tau$, 
if $P(\vec{t})$ is either an internal update or an external update
generated at a time point $\tau'$ no later than $\tau$.  
A node \nodeid sends out a tuple $P(\nodeid', \vec{t})$ if
that tuple was derived by node \nodeid. Because $\nodeid'$ is different
from \nodeid, it is sent over the network.
 A {\em received tuple} is one that comes
from the network (obtained using \rulename{DeQueue}). Finally, an honest node \nodeid runs \prog at time
$\tau$, if at time $\tau$ and the local state of \nodeid at time $\tau$ is the initial
state with an empty table and update queue.

\begin{figure}[t!]
\begin{tabbing}
$\trace \vDash P(\vec{t}) @ (\nodeid,\tau)$  iff 
 $\exists\tau'\leq \tau$,
$\conf$ is the configuration on $\trace$ prior to time $\tau'$,
\\~~\=$ (\nodeid, \tuples, \updates, \prog(\nodeid))\in\conf$, 
 at time $\tau'$,
 $(\nodeid, \tuples, \updates, \prog(\nodeid)) \stepsone
(\nodeid, \tuples', \updates'\circ\updates_\textit{in}, \prog(\nodeid)),\updates_\textit{e}$,
\\\>and  either $P(\vec{t})\in\updates_\textit{in}$
or $P(\vec{t})\in\updates_\textit{e}$
\\
$\trace \vDash \pred{send}(\nodeid, \pred{tp}(P, \nodeid', \vec{t})) @
\tau$  iff 
$\conf$ is the configuration on $\trace$ prior to time $\tau$,
\\~~\=$ (\nodeid, \tuples, \updates, \prog(\nodeid))\in\conf$,
 at time $\tau$,
 $(\nodeid, \tuples, \updates, \prog(\nodeid)) \stepsone \lstate',
\updates_\textit{e}$ 
and 
 $P(@\nodeid', \vec{t})\in\updates_\textit{e}$
\\
$\trace \vDash \pred{recv}(\nodeid, \pred{tp}(P,\vec{t})) @
\tau$  iff   $\exists\tau'\leq \tau$,
$\conf\steps{\tau'}\conf'\in\trace$,  
\\~~\= \queue is the network queue in \conf,
 $P(\vec{t})\in\queue$,
 $(\nodeid, \tuples, \updates, \prog(\nodeid))\in\conf'$ and $P(\vec{t})\in\updates$
\\
$\trace\vDash \pred{honest}(\nodeid, \prog(\nodeid), \tau)$ iff
 at time $\tau$, 
 node \nodeid's local state is (\nodeid, [], [], \prog(\nodeid))
\\[3pt]
$\Gamma \vDash \runby{\prog}{i} : \{i, y_b,y_e\}.\varphi(i, y_b,y_e)$ iff
 Given any trace $\trace$ such that $\trace\vDash \Gamma$,
\\~~\= and at time $\tau_b$, 
 node \nodeid's local state is (\nodeid, [], [], $\prog(\nodeid)$)
\\\> given any time point $\tau_e$ such that  $\tau_e \geq \tau_b$, it
is the case that $\trace\vDash \varphi(\nodeid, \tau_b, \tau_e)$
\end{tabbing}
\caption{Trace-based semantics}
\label{fig:semantics}
\end{figure}

The semantics of invariant assertion states that if a trace $\trace$
contains the execution of \prog by node \nodeid (formally defined as
the node running \prog is one of the nodes in the configuration
$\conf$), then given any time point $\tau_e$ after $\tau_b$, the trace
$\trace$ satisfies $\varphi(\nodeid, \tau_b, \tau_e)$. Here, the
semantic definition requires that the invariant of an honest node
holds in the presence of attackers, because we examine all traces that
include the honest node in their configurations. This means that those
traces can contain arbitrary other nodes, some of which are
malicious.

Our program logic is proven to be sound with regard to the trace
semantics: 

\begin{thm}[Soundness]
\label{thm:soundness}\hfill

\begin{packedenumerate}
\item If $\Sigma;\Gamma\vdash \varphi$, then for all grounding substitution
$\sigma$ for $\Sigma$, given any trace $\trace$, 
$\trace\vDash \Gamma\sigma$ implies $\trace\vDash \varphi\sigma$;
\item If $\Sigma;\Gamma\vdash \runby{\prog}{i} :  \{i, y_b,y_e\}.\varphi(i, y_b,y_e)$,
  then for all grounding substitution
$\sigma$ for $\Sigma$, $\Gamma\sigma\vDash \runby{(\prog)\sigma}{i} :  \{i, y_b,y_e\}. (\varphi(i, y_b,y_e))\sigma$.
\end{packedenumerate}
\end{thm}

\noindent The detailed proof of Theorem~\ref{thm:soundness} can be found in
Section~\ref{app:soundness}. The intuition behind the soundness proof is
that the invariant properties $\varphi_p$ specified for the predicate
$p$ are local properties that will not be affected by the
attacker. For instance, we can specify basic
arithmetic constraints of arguments derived by the honest node and the
existence of base tuples. These invariants can be checked by examining
the program of the honest node and are not affected by how the honest
node interacts with the rest of the network. We never use
any invariant of received tuples, because they could be sent from an
attacker, and the attacker does not need to generate those tuples
following protocols. However, we can use the fact that those received
tuples must have arrived at the honest node; otherwise, the rule will
not fire. In other words, we trust the runtime of an honest node.

\Paragraph{Discussion.}
Our program logic enables us to prove invariant properties that hold even in
adversarial environment. The network trace $\trace$ in Theorem~\ref{thm:soundness} could
involve attacker threads who run arbitrary malicious programs. For example, a
trace may contain attacker threads who keep propagating invalid
route advertisement for a non-existent destination. Properties proved
with our logic, however, still hold in such traces. The key observation here is
that in the rule \rulename{inv}, the correctness of the program property does
not rely on received tuples, which could have been manipulated by malicious
attackers. This guarantee is further validated by our logic semantics and
soundness, where we demand that a proved conclusion should hold in \textit{any}
trace.

Our program logic could possibly prove false program invariants for \Langsec
programs only generating empty network traces. A such example program is as follows:

\[
\begin{array}{l@{~~}l}
\textit{r1}&\pred{p}(@a) \derives \pred{q}(@a).
\\
\textit{r2}&\pred{q}(@a) \derives \pred{p}(@a).
\end{array}
\]
A user could assign $\tfalse$ to both \pred{p} and \pred{q}, and prove the
program invariant with the rule \rulename{inv}. However, this program, when
executing in bottom-up evaluation, produces an empty set of tuples. The
\rulename{inv} rule is still sound in this case as there is no trace that
generates tuples \pred{p} and \pred{q}.
Instead, a \Langsec program should have rules of the form
``\textsf{p} :-'' to generate base tuples. If a false program invariant is given
for such a program, the user is obliged to prove $\vdash \tfalse$ in the logic,
which is impossible. 

\section{Verification Condition Generator}
\label{sec:implementation}
As a step towards automated verification,
  we implement a verification condition generator (\VCG) to
  automatically extract proof obligations from a \Langsec program.
\VCG is implemented in C++ and fully integrated to
 RapidNet~\cite{rapidneturl}, a declarative networking engine for
 compiling \Langsec programs.
We target Coq, but other interactive theorem provers such as Isabelle
HOL are possible.  

Concretely, \VCG generates lemmas
corresponding to the last premise of rule \rulename{Inv}. It takes as
inputs: the abstract syntax tree of a \Langsec program $\mathit{sp}$,
and type annotations $\mathit{tp}$. 
The generated Coq file contains the following:
(1) definitions for types, predicates, and functions;
(2) lemmas for rules in the program;
and (3) axioms based on \rulename{Honest} rule.

\Paragraph{Definition.}
Predicates and functions are
declared before they are used.
Each predicate (tuple) $\mathit{p}$ in the
\Langsec program corresponds to a predicate of the same name in the Coq
file, with two additional arguments: 
a location specifier and a time point.

For example, the generated declaration of the \textit{link} tuple
$\pred{link}(@node, node)$ is the following

\begin{center}
 \pred{Variable} \pred{link}: \pred{node} $\rightarrow$ \pred{node}  $\rightarrow$
  \pred{node} $\rightarrow$ \pred{time}  $\rightarrow$ \pred{Prop}. 
\end{center}
For each user-defined function, 
a data constructor of the same name is generated, unless it corresponds to
a Coq's built-in operator (e.g. list operations). 
The function takes a time point as an additional argument.

\Paragraph{Lemmas.}
For each rule in
a \Langsec program, \VCG generates a lemma
in the form of the last premise in
inference rule \rulename{Inv}
(Figure~\ref{fig:program:logic}).
Rule \textit{sp1} of example program in
Section~\ref{sec:sandlog-syntax}, for
instance, corresponds to the following
lemma:

{\small\textsf{Lemma r1: forall(s:node)(d:node)(c:nat)(p:list node)(t:time),
\\$~~~~~~~~~$\texttt{link s d c s t $\rightarrow$ p = cons (s (cons
  d nil)) $\rightarrow$ p-path s t s d c p t}}.}
\vspace{3pt}

Here, \textit{cons} is Coq's built-in
list appending operation. and \textit{p-path} is the invariant
associated with predicate \textit{path}.

\Paragraph{Axioms.}
For each invariant $\varphi_p$ of a rule head $p$, \VCG produces an axiom
of the form: $\forall i, t, \vec{x}, \pred{Honest}(i)\imp p(\vec{x})@(i, t)
\imp\varphi_p(i, \vec{x})$. 
These axioms are conclusions of the \rulename{Honest} rule after
invariants are verified. Soundness of these axioms is backed by         
Theorem~\ref{thm:soundness}.
Since we always assume that the program starts at time $-\infty$,
the condition that $t>-\infty$ is always true, thus omitted.

\section{Case Studies}
\label{sec:casestudy}

In this section, we investigate two proposed secure routing solutions:
S-BGP (Section~\ref{sec:casestudy:sbgp}) and SCION
(Section~\ref{sec:casestudy:scion}). We encode both solutions in
\Langsec and prove that they preserve route authenticity, a key
property stating that route announcements are trustworthy. Our case
studies not only demonstrate the effectiveness of our program logic,
but provide a formal proof supporting the informal guarantees given by
the solution designers. Interested readers can find \Langsec
specification and formal verification of both solutions online
(\url{http://netdb.cis.upenn.edu/secure_routing/}.)

\subsection{S-BGP}
\label{sec:casestudy:sbgp}

\chen{Generally, the encoding of the protocol and of the property are
  not so clear. The protocol is presented informally in the text, and
  formally in Appendix - some middle ground is missing. For instance,
  how does the axiom for the signature relate to the program; when are
  ‘verify’ and ‘signature’ applied in rules?}
Secure Border Gateway Protocol (S-BGP)~\cite{sbgp} is a comprehensive solution
that aims to eliminate security vulnerabilities of BGP, while maintaining
compatibility with original BGP specifications. S-BGP requires that
each node sign the route information (route attestation) using asymmetric
encryption (e.g. RSA~\cite{RSA})
before advertising the message to its neighbor. The
route information is supposed to include the destination address (represented by an IP
prefix), the known path to the destination, and
the identifier of the neighbor to whom the route information will be sent. The
sender also attaches a signature list to the route information, containing all
signatures received from the previous neighbors.
A node receiving the route attestation would not trust the routing
information unless all signatures inside are properly checked.

\Paragraph{Encoding.}
Figure~\ref{fig:sbgp:encoding} presents our encoding of S-BGP in
\Langsec. The meaning of tuples in the program can be found in
Figure~\ref{fig:sbgp:tuples}.
In rule r1 of Figure~\ref{fig:sbgp:encoding}, when a node \textsf{N} receives an
\pred{advertise} tuple from its neighbor \textsf{Nb}, it generates a \pred{verifyPath}
tuple, which serves as an entry point for recursive signature verification. In
rule 2, \textsf{N} recursively verifies all signatures in \textsf{Osl}, which stands for
``original signature list''. \textsf{Sl} in \pred{verifyPath} is a sub-list of \textsf{Osl},
representing the signatures that have not been checked. When all signatures have
been verified --- this is ensured by ``$f\_size(Sl) == 0$'' in rule 3 --- \textsf{N}
accepts the route and stores the path as a \pred{route} tuple in the local
database. Rule 4 also allows $N$ to generate a \pred{route} tuple storing the
path to its self-owned IP prefixes
(i.e. $\pred{prefix}(@\textsf{N},\textsf{Pfx})$).
Given a specific destination \textsf{Pfx}, in rule 5, $N$ aggregates all \pred{route}
tuples storing paths to \textsf{Pfx}, and computes a \pred{bestPath} tuple for
the shortest path. The \pred{bestPath} is intended to be propagated to
downstream ASes. Before propagation, however, S-BGP requires $N$ to sign the path
information. This is captured in rule 6, where $N$ uses its private key
(i.e. $\pred{privateKeys}(@\textsf{N},\textsf{PriK})$) to generate a signature
based on the selected \pred{bestPath} tuple. Finally, in rule 7, $N$ embeds the
routing information (i.e. \pred{bestPath}) along with its signature
(i.e. \pred{signature}) into a new route advertisement (i.e. \pred{advertise}),
and propagates the message to its neighbors.

\begin{figure*}[th!]
\begin{minipage}[t]{.5\textwidth}
\begin{NDlog}
r1 verifyPath(@N,Nb,Pfx,Pvf,
               Sl,OrigP,Osl) :-
    advertise(@N,Nb,Pfx,RcvP,Sl),
    link(@N,Nb),
    Pvf := f_prepend(N,RcvP),
    OrigP := Pvf, 
    Osl := Sl,
    f_member(RcvP,N) == 0,
    Nb == f_first(RcvP).
\end{NDlog}
\vspace{4pt}

\begin{NDlog}
r2 verifyPath(@N,Nb,Pfx,PTemp,
                Sl1,OrigP,Osl) :-
    verifyPath(@N,Nb,Pfx,Pvf,
                Sl,OrigP,Osl),
    publicKeys(@N,Nd,PubK),
    f_size(Sl) > 0, 
    f_size(Pvf) > 1,
    PTemp := f_removeFirst(Pvf), 
    Nd := f_first(PTemp), 
    SigM := f_first(Sl), 
    MsgV := f_prepend(Pfx,Pvf),
    f_verify(MsgV,SigM,PubK) == 1,
    Sl1 := f_removeFirst(Sl).
\end{NDlog}
\vspace{4pt}
\begin{NDlog}
r3 route(@N,Pfx,C,OrigP,Osl) :-
    verifyP(@N,Nd,Pfx,Pvf,
             Sl,OrigP,Osl),
    f_size(Sl) == 0, 
    f_size(Pvf) == 1, 
    C:= f_size(OrigP) - 1.
\end{NDlog}
\end{minipage}
\vspace{8pt}

\begin{minipage}[t]{.5\textwidth}
\begin{NDlog}
r4 route(@N,Pfx,C,P,Sl) :-
    prefixs(@N,Pfx), 
    List := f_empty(), 
    C := 0,
    P := f_prepend(N,List), 
    Sl := f_empty().
\end{NDlog}
\vspace{4pt}
\begin{NDlog}
r5 bestRoute(@N,Pfx,a_MIN<C>,P,Sl) :-
    route(@N,Pfx,C,P,Sl).
\end{NDlog}
\vspace{4pt}
\begin{NDlog}
r6 signature(@N,Msg,Sig) :-
    bestRoute(@N,Pfx,C,BestP,Sl),
    privateKeys(@N,PriK),
    link(@N,Nb), 
    Pts := f_prepend(Nb,BestP),
    Msg := f_prepend(Pfx,Pts),
    Sig := f_sign(Msg,PriK).
\end{NDlog}
\vspace{4pt}
\begin{NDlog}
r7 advertise(@Nb,N,Pfx,BestP,NewSl) :-
    bestRoute(@N,Pfx,C,BestP,Sl),
    link(@N,Nb),
    Pts := f_prepend(Nb,BestP),
    Msg == f_prepend(Pfx,Pts),
    signature(@N,Msg,Sig),
    NewSl := f_prepend(Sig,Sl).
\end{NDlog}
\end{minipage}
\caption{S-BGP encoding}
\label{fig:sbgp:encoding}
\end{figure*}

\begin{figure*}[t!]
\begin{tabular}{ll}
$\pred{link}(@n, n')$ & there is a link between $n$ and $n'$. \\ 
$\pred{route}(@n, d, c,p,sl)$ & $p$ is a path to $d$ with cost $c$. 
\\
 ~& $sl$ is the signature list associated with $p$. \\
$\pred{prefix}(@n, d)$ & $n$ owns prefix (IP addresses) $d$. 
\\
$\pred{bestRoute}(@n, d, c, p, sl)$ & $p$ is the best path to $d$ with
cost $c$. 
\\ ~ &$sl$ is the signature list associated with $p$. 
\\

$\pred{verifyPath}(@n, n', d, p, sl,$ 
 & a path $p$ to a destination $d$ is verified against signature list $sl$.
\\~~~~~~~~~~~~~ $\textit{pOrig}, \textit{sOrig})$ & $p$ is a sub-path of \textit{pOrig}, and $s$ is a sub-list of \textit{sOrig}.
\\
$\pred{signature}(@n, m, s)$ & $n$ creates a signature $s$ of 
message $m$ with private key.
\\
$\pred{advertise}(@n', n, d, p, sl)$ & $n$ advertises path $p$ to neighbor $n'$ with signature list $sl$.
\end{tabular}
\caption{Tuples for $\sbgp$\reviewer{This is the first time you actually show some crypto;you might want to add some forward references to keep some people
from giving up before getting to it.}}
\label{fig:sbgp:tuples}
\end{figure*}

\Paragraph{Property specification.}
Route authenticity of S-BGP ensures that no route announcement can be
tampered with by an attacker.  In other words, it requires that
any route announcement \textit{accepted} by a node is authentic.
We encode it as
$\varphi_\textit{auth1}$
below. 

\vspace{3pt}
\fbox{%
  \parbox{0.9\textwidth}{%
  \noindent$\varphi_\textit{auth1}$ =$\forall n, m, t,d, p, sl, \\
  ~~~~~~~~~~~~~~~~
  \pred{Honest}(n)
  \conjunc \pred{advertise}(m, n, d, p, sl)@(n, t)$
  $\imp \pred{goodPath}(t, d, p)$
  }
}
\vspace{3pt}

\noindent $\varphi_\textit{auth1}$ is a general topology-independent security
property. It asserts that whenever an honest node
$n$, denoted as $\pred{Honest}(n)$, sends
out an \pred{advertise} tuple to its neighbor $m$, the property $\pred{goodPath}(t, d, p)$
holds. $\pred{Honest}(n)$ means that $n$ runs S-BGP and
$n$'s private key is not compromised. Formally:

\vspace{3pt}
\fbox{%
  \parbox{0.5\textwidth}{%
    $\pred{Honest}(n)\defeq \pred{honest}(n, \sbgp(n),
    -\infty)$. 
}
}
\vspace{3pt}

\noindent Here, the starting time
is set to be the earliest possible time point. 
\Langsec's semantics allows a node to begin execution at any time
after the specified starting time, so using $-\infty$ gives us the most
flexibility.
$\pred{goodPath}(t, d, p)$ is recursively defined in Figure~\ref{fig:goodpath},
which asserts that all links in the path $p$ towards the destination $d$ exist no
later than $t$. 
Each link ($m$, $n$) is represented by two tuples: \pred{link(@n, m)} and
\pred{link(@m, n)}. These two tuples reside on two endpoints respectively.
\reviewer{Another thing I found strange is that axiom Asig uses the Honest
predicate...So if instead of your program, I would use your framework to analyze the protocol ``A sends her signing key to B, and B uses it to sign a message and send
it to C'', you would basically be assuming False and accept the protocol?  Wouldn't it be better to reject such clearly broken protocols?}

\begin{figure}[t!]
\begin{mathpar}
\mprset{flushleft}
\inferrule*{ 
\pred{Honest}(n) \imp
\exists t', t'\leq t \conjunc \pred{prefix}(n, d)@(n, t') \\
}{
\pred{goodPath}(t, d, n::\f{nil})
}

\figspace

\inferrule*{
\pred{Honest}(n)\imp \exists t', t'\leq t \conjunc \pred{link}(n,
 n')@(n, t') 
\\
\pred{goodPath}(t, d, n::\f{nil})
}{
\pred{goodPath}(t, d, n'::n::\f{nil})
}

\figspace

\inferrule*{
\pred{Honest}(n)\imp \exists t', t'\leq t \conjunc \pred{link}(n,
 n')@(n, t') \conjunc
\exists t'', t''\leq t\conjunc \pred{link}(n, n'')@(n, t'')
\\ 
\pred{goodPath}(t, d, n::n''::\f{p''})
}{
\pred{goodPath}(t, d, n'::n::n''::\f{p''})
}
\end{mathpar}
\caption{Definitions of \pred{goodPath}}
\label{fig:goodpath}
\end{figure}

To be more specific, the definition of $\pred{goodPath}(t, d, p)$ involves three
cases (Figure~\ref{fig:goodpath}). The base case is when $p$ contains only one node. We require that $d$
be one of the prefixes owned by $n$ (i.e., the \pred{prefix} tuple is
derivable).  When $p$ has two nodes $n'$ and $n$, we require that the link
from $n$ to $n'$ exist from $n$'s perspective, assuming that $n$ is
honest, but impose no constraint on $n'$'s database, because $n'$ has not
received the advertisement. The last case is when the
  length of $p$ is larger than two; we check that both links (from $n$ to
$n'$ and from $n$ to $n''$) exist from $n$'s perspective, assuming $n$
is honest. In the last two rules, we also recursively check that the
subpath also satisfies \pred{goodPath}.

\pred{goodPath} can serve as a template for a number of useful properties.
For example, by substituting \pred{link}($n$,$n''$) with
\pred{announce\_link}($n$, $n''$), we are able to express whether a node is
willing to let its neighbor know of that link. We can also require each
subpath be authorized by the sender.

\Paragraph{Axiom of signature.}
To use the authenticity property of signatures in the proof of
$\varphi_\textit{auth1}$, we include the following axiom $A_\textit{sig}$ in the
logical context $\Gamma$. This axiom states that if a signature $s$ is verified by the
public key of a node $n'$, and $n'$ is honest, then $n'$ must have generated a
\pred{signature} tuple.  
Predicate $\pred{verify}(m, s, k)@(n,t)$ means
that node $n$ verifies, using key $k$ at time $t$, that $s$ is a valid signature
of message $m$.

\vspace{3pt}

\fbox{%
  \parbox{0.9\textwidth}{%
 \begin{tabbing}
 $A_\textit{sig}$ = \= $\forall m, s,k,n,n',t, \pred{verify}(m, s, k)@(n, t) \conjunc $
 $\pred{publicKeys}(n, n',
   k)@(n, t) \conjunc $ 
 \\\>~~~$ \pred{Honest}(n') \imp\exists t', t'< t\conjunc \pred{signature}(n', m, s)@(n', t')$
 \end{tabbing}
}
}
\bigskip

\Paragraph{Verification.}
Our goal is to prove that $\varphi_\textit{auth1}$ is an
  invariant property that holds on all possible execution traces. However,
  directly proving $\varphi_\textit{auth1}$ is hard, as it involves verification
  over all the traces. Instead, we take the indirect approach of using our program logic
  to prove a program invariant, which is stronger than $\varphi_\textit{auth1}$,
  and, more importantly, whose validity implies the validity of
  $\varphi_\textit{auth1}$. To be concrete, we show that \sbgp has the following invariant property $\varphi_{I}$:

\vspace{3pt}
\fbox{%
  \parbox{0.55\textwidth}{%
\vspace{6pt}
(a) $\cdot; \cdot\vdash \runby{\prog_\textit{sbgp}}{i}: \{i, y_b, y_e\}.\varphi_{I}(i, y_b, y_e)$
\vspace{6pt}
}
}
\vspace{3pt}

where $\varphi_{I}$ is defined as:

\vspace{3pt}
\fbox{%
\parbox{0.93\textwidth}{%
$\varphi_{I}(i, y_b, y_e) = 
  \bigwedge_{p\in\headof{\sbgp}}\forall t\hspace{3pt}\vec{x}, y_b\leq t<y_e\conjunc
 p(\vec{x})@(i, t)
  \imp \varphi_{p}(i, t, \vec{x})$
}
}
\vspace{3pt}
\limin{what are all the $\varphi_p$s? The reviewer asks us to be very
  clear on the case study. I think it is a good idea to use a table to
  list all the $\varphi_p$s.}

\noindent Every $\varphi_p$ in $\varphi_{I}$ denotes the invariant property
associated with each head tuple in \sbgp, and 
needs to be specified by the user. Table~\ref{tab:inv} gives the invariants
associated with all head tuples in the program. Especially, the invariant associated
with the \pred{advertise} tuple (\pred{goodPath}) is the same as
the conclusion of $\varphi_\textit{auth1}$. 

We prove (a) using the \rulename{inv} rule in
Section~\ref{sec:verification}, by showing that all the premises
  hold. The \rulename{inv} rule has two types of premises: (1) Premises that ensure
  each rule of the program maintains the invariant of its rule head; and (2) Premises
that ensure all invariants for head tuples are closed under trace extension. Premises of
the second type are guaranteed through manual inspection of all the invariants,
thus omitted in the formal proof. As for premises of the first type, since
\sbgp has seven rules, this corresponds to seven premises to be proved. For
example, the premise corresponding to rule 2 is represented by
($a_0$), shown below. 

\begin{table}[t!]

\centering
  \begin{tabular}{|c|l|l|}
  \hline
  \textbf{Rule} & \textbf{Head Tuple} & \textbf{Invariant} \\ \hline
  r1,r2 & 
  \pred{verifyPath}(\textsf{N,Nb,Pfx,Pvf,} & 
  $\exists l, \textsf{Osl} = l {++} \textsf{Pvf} \conjunc$ \\

  &
  ~~~~~ \textsf{Sl,OrigP,Osl})@(\textsf{N,t}) & 
  $(\pred{goodPath}(\textsf{t,Pfx,Pvf}) \imp
  \pred{goodPath}(\textsf{t,Pfx,Osl}))$\\ \hline

  r3,r4 & 
  \pred{route}(\textsf{N,Pfx,C,OrigP,Osl})@(\textsf{N,t}) & 
  \pred{goodPath}(\textsf{t,Pfx,OrigP})\\ \hline

  r5 &
  \pred{bestRoute}(\textsf{N,Pfx,C,P,Sl})@(\textsf{N,t}) &
  \pred{goodPath}(\textsf{t,Pfx,OrigP}) \\ \hline

  r6 & 
  \pred{signature}(\textsf{N,Msg,Sig})@(\textsf{N,t}) & 
  $\exists \textsf{p},\textsf{m},\textsf{pfx}, \textsf{Msg} =
  \textsf{pfx}::\textsf{nei}::\textsf{p}$ \\ \hline

  r7 & 
  \pred{advertise}(\textsf{Nb,N,Pfx,BestP,NewSl}) & 
  \pred{goodPath}(\textsf{t,Pfx,Nb::BestP}) \\
  \hline
  \end{tabular}
  \caption{Tuple invariants in $\varphi_I$ for S-BGP route authenticity}
  \label{tab:inv}
\vspace{-6pt}
\end{table}

\vspace{3pt}
\fbox{%
  \parbox{0.9\textwidth}{%
 (a$_0$) $\cdot;\cdot\vdash$
 $\forall \textsf{N}, \forall \textsf{Nb}, \forall \textsf{Pfx},\forall \textsf{Pvf},\forall \textsf{Sl},\forall \textsf{Sl1},\forall
\textsf{OrigP},\forall \textsf{Osl},\forall \textsf{t},\forall \textsf{Nd},\\
~~~~~~~~~~~~~~\forall \textsf{PubK},\forall \textsf{m},\forall \textsf{p},\forall \textsf{SigM},\forall \textsf{MsgV},\forall \textsf{PTemp},\forall \textsf{Osl},\\
~~~~~~~~~~~~~~\pred{verifyPath}(\textsf{N,Nb,Pfx,Pvf,SL,OrigP,Osl})@(\textsf{N,t})
\conjunc \\
~~~~~~~~~~~~~~\exists l, \textsf{Osl} = l {++} \textsf{Sl} \conjunc \\
~~~~~~~~~~~~~~(\pred{goodPath}(\textsf{t,Pfx,Pvf}) \imp \pred{goodPath}(\textsf{t,Pfx,Osl}))
\conjunc \\
~~~~~~~~~~~~~~\pred{publicKeys}(\textsf{N,Nd,PubK})@(\textsf{N,t}) \conjunc \\
~~~~~~~~~~~~~~\pred{length}(\textsf{Sl}) > 0 \conjunc \\
~~~~~~~~~~~~~~\pred{length}(\textsf{Pvf}) > 0 \conjunc \\
~~~~~~~~~~~~~~\textsf{Pvf} = \textsf{m :: Nd :: p} \conjunc \\
~~~~~~~~~~~~~~\textsf{PTemp} = \textsf{Nd :: p} \conjunc \\
~~~~~~~~~~~~~~\textsf{Sl} = \textsf{SigM :: Sl1} \conjunc \\
~~~~~~~~~~~~~~\textsf{MsgV} = \textsf{Pfx :: Pvf} \conjunc \\
~~~~~~~~~~~~~~\pred{verify}(\textsf{MsgV,SigM,PubK})@(\textsf{N,t}) \imp\\
~~~~~~~~~~~~~~~~~~(\exists l, \textsf{Osl} = l {++} \textsf{Sl1} \conjunc \\
~~~~~~~~~~~~~~~~~~(\pred{goodPath}(\textsf{t,Pfx,PTemp}) \imp
\pred{goodPath}(\textsf{t,Pfx,Osl})))$

}
}
\bigskip

\noindent Here, \textit{$(\exists l, \textsf{Osl} = l {++} \textsf{Sl1} \conjunc
(\pred{goodPath}(\textsf{t,Pfx,PTemp}) \imp
\pred{goodPath}(\textsf{t,Pfx,Osl})))$} is the invariant of rule 2's head tuple
$\pred{verifyPath}$ (Figure~\ref{tab:inv}). Other rule-related premises are
constructed in a similar way. We prove all the premises in Coq, thus proving
($a$).

After (a) is proved, by applying the \rulename{Honest} rule, we can deduce
$\varphi=\forall n\hspace{3pt} t, \pred{Honest}(n) \imp \varphi_I(n, t,
-\infty)$.  $\varphi_I$ can then be injected into the assumptions ($\Gamma$) by
\VCG (as do $\varphi_{I1}$) and is safe to be used as theorem in proving other properties.
Finally, $\varphi_\textit{auth1}$ is proved by discharging $\varphi \imp
\varphi_\textit{auth1}$ in Coq with standard elimination rules. 

\Paragraph{Proof details.} Among the others, the premise corresponding to rule 2 in the program
turns out to be the most challenging one, as it involves recursion and signature
verification.
Recursion in rule 2 makes it hard to find the proper invariant specification for
the head tuple \pred{verifyPath}, as the invariant needs to maintain correctness
for both the head tuple and the body tuple, which have different arguments.

In our specification, we specify the invariant in a way that reversely verify the signature
list by checking the signature for the longest path first.
More concretely, we use an implication, stating that if the path to be verified
satisfies the invariant \pred{goodPath}, then the entire path satisfies the
invariant \pred{goodPath} (Table~\ref{tab:inv}).

Another challenge in proving the invariant for \pred{verifyPath} is to reason
about the existence of $link$ tuples
at the previous nodes. We solve the problem in two steps:
(1) we prove a stronger auxiliary program invariant ($a_1$), which
asserts the existence of the local $link$ tuple when a node signs the path
information.
(2) we then use the axiom $A_\textit{sig}$ to allow a node who verifies a signature
to assure the existence of the $link$ tuple at the remote node who signs the
signature. 

More concretely, ($a_1$) is defined as:

\vspace{3pt}
\fbox{%
\parbox{0.45\textwidth}{%
(a$_1$) $\cdot; \cdot\vdash \runby{\prog_\textit{sbgp}}{i}:
\{i, y_b, y_e\}.\varphi_{I1}$ 
}
}
\bigskip

\noindent In (a$_1$), all head tuples \textit{p} other than \textit{signature} and
\textit{advertise} takes on the same invariant $\varphi_\textit{link1}(p, n,d, t)$:

\vspace{3pt}
\fbox{%
  \parbox{0.9\textwidth}{%
$\varphi_\textit{link1}(p, n, d, t) = \exists p',\\
~~~~~~~~~~~~~~~~~~~~~~~~~  p = n :: p' \conjunc$ 
$ (p' = \pred{nil} \imp \pred{prefix}(n,
  d)@(n, t))$ 
$\conjunc \\
~~~~~~~~~~~~~~~~~~~~~~~~~ \forall p'', m', p' = m' :: p'' \imp \pred{link}(n, m')@(n, t)$
}
}
\bigskip

\noindent It states that node $n$ is the first element in path $p$, and the
$link$ tuple from $n$ to its neighbor in $p$ exists in $n$'s database.

For \textit{signature} and \textit{advertise}, we introduce another property:

\vspace{3pt}
\fbox{%
  \parbox{0.9\textwidth}{%
$\varphi_\textit{link2}(p, n, d, n', t)$= 
 $\pred{link}(n, n')@(n,t)\conjunc \\
~~~~~~~~~~~~~~~~~~~~~~~~~ \exists p',$ $p = n :: p' \conjunc (p' = \pred{nil} \imp \pred{prefix}(n,
  d)@(n, t))$ 
$\conjunc \\ 
~~~~~~~~~~~~~~~~~~~~~~~~~ \forall p'', m', p' = m' :: p'' \imp \pred{link}(n, m')@(n, t)$
}
}
\bigskip

\noindent $\varphi_\textit{link2}(p, n, d, n', t)$ extends $\varphi_\textit{link1}(p, n,
d, t) $ by including the receiving node $n'$ as an argument, asserting that the
link between $n$ and $n'$ also exists.
And the invariants of \textit{signature} and \textit{advertise} are:

\vspace{3pt}
\fbox{%
  \parbox{0.85\textwidth}{%
$\varphi_\textit{signature}(i,t, n, m, s)=\exists n', d, m = d::n'::p \conjunc
\varphi_\textit{link2}(p, n, d, n', t)$

$\varphi_\textit{advertise}(i,t,n',n,d,p,\textit{sl})=
\varphi_\textit{link2}(p,n,d,n',t)$
}
}
\bigskip

\noindent We prove ($a_1$) using the \rulename{Inv} rule.

Then, by applying \rulename{Honest} rule to (a$_1$) and only keeping the clause
in $\varphi_{I2}$ related to \pred{signature}, we derive the following:

\vspace{3pt}
\fbox{%
  \parbox{0.9\textwidth}{%
 (a$_2$) $\cdot;\cdot\vdash$
 $\forall n, \forall t, \forall m,\\
~~~~~~~~~~~~~~\pred{Honest}(n)\conjunc\pred{signature}(n,
  m, s)@(n, t) \imp \\
~~~~~~~~~~~~~~\exists n',d,p  m = d::n'::p \conjunc\varphi_\textit{link2}(p, n, d, n', t)$
}
}
\bigskip

\noindent($a_2$) connects an honest node's signature to the existence of
related link tuples at a previous node in the path $p$.

Next, we use ($a_2$) along with $A_\textit{sig}$ to prove (a$_0$). Applying
$A_\textit{sig}$ to tuples \pred{publicKeys} and \pred{verify} in ($a_0$), we
can get:

\vspace{3pt}
\fbox{%
  \parbox{0.9\textwidth}{%
 (a$_3$) $\cdot;\cdot\vdash$
 $\forall \textsf{Nd}, \forall \textsf{MsgV}, \forall \textsf{SigM}, \forall \textsf{t},\\
~~~~~~~~~~~~~~\pred{Honest}(\textsf{Nd}) \imp\exists \textsf{t'}, \textsf{t'} <
\textsf{t} \conjunc \pred{signature}(\textsf{Nd, MsgV,SigM})@(\textsf{Nd,t'})$
}
}
\bigskip

\noindent We further apply ($a_2$) to ($a_3$) to obtain:

\vspace{3pt}
\fbox{%
  \parbox{0.9\textwidth}{%
 (a$_4$) $\cdot;\cdot\vdash$
 $\forall \textsf{Nd}, \forall \textsf{t}, \forall \textsf{MsgV},\\
~~~~~~~~~~~~~~\exists n',d,p, \textsf{Nd} = d::n'::p \conjunc\varphi_\textit{link2}(p, \textsf{Nd}, d, n', \textsf{t})$
}
}
\bigskip

\noindent Combining (a$_4$) and the assumptions in (a$_0$), we are able to prove the conclusion
of (a$_0$). Other premises can be proved similarly. For non-recursive rules, the
premises for them are straightforward. The detailed proof can be found online.

\Paragraph{Discussion.} 
$\varphi_\textit{auth1}$ is a general template for proving different kinds of
route authenticity. For example, 
S-BGP satisfies a stronger property that guarantees authentication of each
subpath in a given path $p$. The property, called $\pred{goodPath2}(t, d, p)$,
is defined in Figure~\ref{fig:goodpath2}. The meaning of the
variables remains the same as before.

\begin{figure}[t!]
\begin{mathpar}
\mprset{flushleft}
\inferrule*{ 
\pred{Honest}(n) \imp
\exists t', t'\leq t \conjunc \pred{prefix}(n, d)@(n, t') \\ 
}{
\pred{goodPath2}(t, d, n::\f{nil})
}

\inferrule*{
\pred{Honest}(n)\imp \exists t', c, \f{s}, t'\leq t \conjunc \pred{link}(n,
 n')@(n, t') \conjunc\pred{route}(n,d,c,n::\f{nil},sl)@(n,t')
\\\\
\pred{goodPath2}(t, d, n::\f{nil})
}{
\pred{goodPath2}(t, d, n'::n::\f{nil})
}

\inferrule*{
\pred{Honest}(n)\imp \exists t', t'\leq t \conjunc \pred{link}(n,
 n')@(n, t') \conjunc
\\~~~~~~~~~~~~~\exists t'', c, \f{s}, t''\leq t\conjunc \pred{link}(n, n'')@(n,
t'')\conjunc
\\~~~~~~~~~~~~~\pred{route}(n,d,c,n::n''::p'',sl)@(n,t')
\\\\
\pred{goodPath2}(t, d, n::n''::\f{p''})
}{
\pred{goodPath2}(t, d, n'::n::n''::\f{p''})
}
\end{mathpar}
\caption{Definitions of \pred{goodPath2}}
\label{fig:goodpath2}
\end{figure}

Compared with \pred{goodPath}, the
last two rules of \pred{goodPath2} additionally assert the existence of a route tuple.
The predicate $\pred{route}(n,d,c,n::p',sl)@(n,t')$ states that node
$n$ generates a route tuple for path $n::p'$ at time $t'$, and that
$sl$ is the signature list that authenticates the path $n::p'$.  This
property ensures that an attacker cannot use $n$'s route advertisement
for another path $p'$, which happens to share the two direct links of $n$.
More specifically, given $p = n1::n::n2::p1$ and $p' = n1::n::n2::p2$, with $\mathit{p1\neq
p2}$, an attacker could not replace $p$ with $p'$ without being detected.
However, a protocol that only requires a node $n$ to sign the links to its direct
neighbors would be vulnerable to such attack.

\subsection{SCION}
\label{sec:casestudy:scion}

SCION~\cite{scion} is a clean-slate design of Internet routing
architecture that offers more flexible route selection and failure
isolation along with route authenticity. Our case study focuses on the
routing mechanism proposed by SCION. We only provide high-level
explanation of SCION. Detailed encoding can be found under the
following link (\url{http://netdb.cis.upenn.edu/secure_routing/}).

\begin{figure}[th!]  
\limin{trim white space around the pdf figure}
\centering 
\includegraphics[scale=0.35]{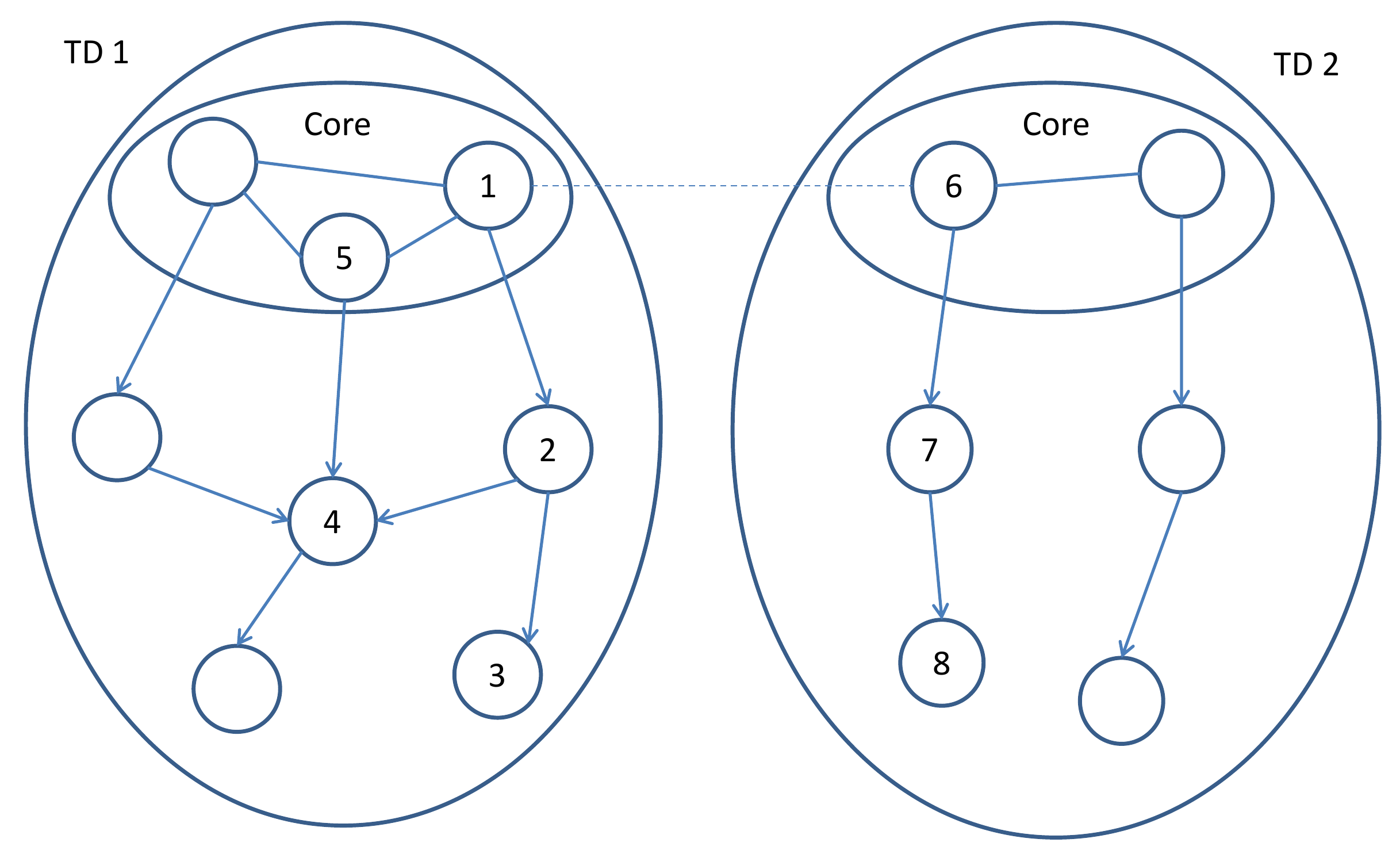}
 \caption{An example deployment of SCION}
 \label{fig:scion}
\end{figure}

In SCION, Autonomous Domains (AD) --- a concept similar to
Autonomous Systems (AS) in BGP --- are grouped into different Trust Domains
(TD). Inside each Trust Domain, top-tier ISP's are selected as the TD
core, which provide routing service inside and across
the border of TD. Figure~\ref{fig:scion} presents an example deployment of SCION
with two TD's. Each AD can communicate with its neighbors. The direction
of direct edges represents provider-customer relationship in routing;
the arrow goes from a provider to its customer. 

\noindent To initiate the routing process, a TD core periodically generates a
path construction announcement, called a beacon, to all its customer ADs. Each
non-core AD, upon receiving a beacon, 
(1) verifies the information inside the beacon, 
(2) attaches itself to the path inside the received beacon to construct a new
beacon, and
(3) forwards the new beacon to its customer ADs.
Each beacon represents a path towards the TD core (e.g. path ``1-2-3'' in Figure~\ref{fig:scion}).
After receiving $k$ beacons , an downstream AD selects $m$ paths out of $k$ and
uploads 
them to the TD core, thus finishing path construction ($k$ and $m$ can be set by
the administrator).  When later an
AD $n$ intends to send a packet to another AD $n'$, it first queries the TD core
for the paths that $n'$ has uploaded, and then constructs a forwarding
path combining its own path to the TD core with the query result. For example,
in Figure~\ref{fig:scion}, when node 4 wants to communicate with node 3, it
would query from the TD core for path ``1-2-3'', and combine it with its own
path to the TD core (i.e. ``4-5''), to get the desired path ``4-5-1-2-3''. 

In Table~\ref{tab:scion}, we summarize the \Langsec encoding of the path construction phase in SCION.
Definitions of important tuples can be found in
Figure~\ref{fig:scion:tuples}. 
The path construction beacon plays an important role in SCION routing mechanism.
A beacon is composed of four fields: an interface field, a time field, an
opaque field and a signature.
The interface field in SCION is identical to the announced path in S-BGP. An
interface field contains a list of AD identifiers representing the routing
path. As its name suggests, the interface field also includes each AD's
interfaces to direct neighbors in the path --- SCION calls the
interface to an AD's provider as \emph{ingress} and the one to a customer as
\emph{egress}. 
Each AD attaches his own identifier along with its \emph{ingress} and
\emph{egress} to the end of the received interface field, generating the new interface field.
For example, in Figure~\ref{fig:scion}, assume the ingress interface of AD 2
against AD1 is ``a'', and the egress interface of AD 2 against AD 4 is
``b''. Given an interface field \{c::1\} from AD 1 --- c represents the egress
interface of AD 1 against AD 2 --- the newly generated interface field at AD 2
targeting AD 4 will be \{c::1::a::b::2\}.

The time field is a list of time stamps which record the arrival time of the
beacon at each AD.  The opaque field
adds a message authentication code (MAC) on each AD's \emph{ingress} and \emph{egress}
fields using the AD's private key, for the purpose of path authentication during data
forwarding. 
The final part is called the signature list. Each AD constructs a signature by
signing the above threes fields (i.e. the interface field, the opaque field and
the time field) along with the signature received from preceding ADs. The newly
generated signature is appended to the end of the signature list.

\begin{table}[htbp!]

\centering
  \begin{tabular}{|c|l|l|}
  \hline
  \textbf{Rule} &
  \multicolumn{1}{|c|}{\textbf{Summary}} &
  \multicolumn{1}{|c|}{\textbf{Head Tuple}} \\ \hline
  \textbf{b1:} & TD core generates a signature. &
  $\pred{signature}(@\mathit{core}, \mathit{info}, \mathit{sig}, \mathit{time})$ \\ \hline
  \textbf{b2:} & TD core signs beacon global information. &
  $\pred{signature}(@\mathit{core}, \mathit{info}, \mathit{sig}, \mathit{time})$ \\ \hline  
  \textbf{b3:} & TD core initiates an opaque field. &
  $\pred{mac}(@core, \mathit{info}, \mathit{hash})$ \\ \hline 
  \textbf{b4:} & TD core initiates global info of beacon. &
  $\pred{beaconPrep}(@\mathit{core}, \mathit{glb}, \mathit{sigG}, \mathit{time})$ \\ \hline
  \textbf{b5:} & TD core sends a new beacon to neighbor. &
  $\pred{beaconIni}(@\mathit{nei}, \mathit{core}, \mathit{td}, \mathit{itf},$ \\
  & & $~~~~~~~~~~~~~~\mathit{tl}, \mathit{ol}, \mathit{sl}, \mathit{sigG})$  \\ \hline
  \textbf{b6:} & AD receives a beacon from TD core. &
  $\pred{beaconRev}(@\mathit{ad}, \mathit{td}, \mathit{td}, \mathit{itf}, $\\
  & & $~~~~~~~~~~~~~~tl, \mathit{ol}, \mathit{ing}, \mathit{sigG})$  \\ \hline
  \textbf{b7:} & AD receives a beacon from non-core AD. &
  $\pred{beaconRev}(@\mathit{ad}, \mathit{td}, \mathit{ing}, \mathit{itf},$ \\
  & & $~~~~~~~~~~~~~~tl, \mathit{ol}, \mathit{sl}, \mathit{sigG})$  \\ \hline
  \textbf{b8:} & AD verifies global information.&
  $\pred{beaconToVeri}(@\mathit{ad}, \mathit{td}, \mathit{itf}, \mathit{l}, \mathit{ol},$ \\
  & & $~~~~~~~~~~~~~~~~~~\mathit{sl, sigG, itfv, pos})$  \\ \hline
  \textbf{b9:} & AD recursively verifies signatures. &
  $\pred{beaconToVeri}(@\mathit{ad, td, itf,tl, ol,}$ \\
  & & $~~~~~~~~~~~~~~~~~~\mathit{sl, sigG, itfv, pos})$  \\ \hline
  \textbf{b10:} & AD validates a beacon. &
  $\pred{verifiedBeacon}(@\mathit{ad, td, ing, itf,}$ \\
  & & $~~~~~~~~~~~~~~\mathit{tl, ol, sl, sigG})$  \\ \hline
  \textbf{b11:} & AD creates signature for new beacon. &
  $\pred{signature}(@\mathit{ad, info, sig, time})$ \\ \hline
  \textbf{b12:} & AD initiates opaque field for new beacon. &
  $\pred{mac}(@\mathit{ad, info, hash})$ \\ \hline 
  \textbf{b13:} & AD sends the new beacon to neighbor. &
  $\pred{beaconFwd}(@\mathit{nei, ad, td, itf,}$ \\ 
  & & $~~~~~~~~~~~~~~\mathit{tl, ol, sl, sigG})$  \\ \hline \hline
  \textbf{pc1:} & AD extracts path information. &
  $\pred{upPath}(@\mathit{ad, td, itf, ol, tl})$ \\ \hline
  \textbf{pc2:} & AD initiates path upload. &
  $\pred{pathUpload}(@\mathit{nei, ad, src, core,}$ \\
  & & $~~~~~~~~~~~~~~\mathit{itf, ol, op, pos})$  \\ \hline
  \textbf{pc3:} & AD sends path to upstream neighbor. &
  $\pred{pathUpload}(@\mathit{nei, ad, src, core,}$ \\
  & & $~~~~~~~~~~~~~~\mathit{itf, ol, op, pos})$  \\ \hline
  \textbf{pc4:} & TD core stores received path.&
  $\pred{downPath}(@\mathit{core, src, itf, op})$ \\ \hline

  \end{tabular}
  \vspace{1mm}
  \caption{\Langsec encoding of path construction in SCION}
  \label{tab:scion}
\vspace{-6pt}
\end{table}

\begin{figure*}
\begin{tabular}{l@{~~~}l}

$\pred{coreTD}(@n,c,\f{td},\f{ctf})$ & $c$ is the core of TD $\f{td}$ with certificate $\f{ctf}$ attesting to that fact \\
$\pred{provider}(@n,m,\f{ig})$ & $m$ is $n$'s provider, with traffic into $n$ through interface $\f{ig}$ \\
$\pred{customer}(@n,m,\f{eg})$ & $m$ is $n$'s customer, with traffic out of $n$ through interface $\f{eg}$ \\
$\pred{beaconIni}(@m,n,\f{td},$ 
& $\f{itf}$, containing a path, is initialized by $n$ and sent to
$m$. \\
~~~~~~~
$\f{itf},\f{tl},\f{ol},\f{sl},\f{sg})$
 & $\f{tl}$ is a list of time stamps, \\
& $\f{ol}$ is a list of opaque fields, whose meaning is not relevant
here. \\
& $\f{sl}$ is list of signatures for route attestation. \\
& $\f{sg}$ is a signature for certain global information, 
\\ &which is not
relevant here.\\
$\pred{verifiedBeacon}(@n,td,$ & $itf$ is the stored interface fields
from $n$ to the TD core in $td$. \\
~~~~~~~~
$itf,tl,ol,sl,sg)$
 & Rest of the fields have the same meaning as those
in \pred{beaconIni} \\
$\pred{beaconFwd}(@m,n,\f{td},$ 
& $\f{itf}$ is forwarded to $m$ with corresponding signature list
$\f{sl}$ \\
~~~~~~~
$\f{itf},\f{tl},\f{ol},\f{sl},\f{sg})$.
& Rest of the fields have the same meaning as those
\\ & in \pred{beaconIni} \\
$\pred{upPath}(@n,\f{td},\f{itf},$
& $\f{opqU}$ is a list of opaque fields indicating a path. \\
~~~~~~~
$\f{opqU},\f{tl})$ & Rest of the fields have the same meaning as those in $\pred{beaconIni}$. \\ 
$\pred{pathUpload}(@m,n,$ & $\f{src}$ is the node (AD) who initiated the path upload process. \\
~~~~~~~
$\f{src},\f{c},\f{itf},\f{opqD},$ & $\f{c}$ is TD core of an implicit TD.\\
~~~~~~~
$\f{opqU},\f{pt})$ & $\f{opqD}$ is the opaque fields uploaded.\\
~~~~~~~
& $\f{pt}$ indicates the next opaque field in $\f{opqU}$ to be checked. \\
& $\f{itf}$ and $\f{opqU}$ have the same meaning as those in $\pred{upPath}$.

\end{tabular}
\caption{Tuples for SCION}
\label{fig:scion:tuples}
\end{figure*}

SCION also satisfies similar route authenticity properties as
S-BGP. Each path in SCION is composed of two parts: a path from the sender to
the TD core (called ``up path'') and a path from the TD core to the receiver
(called ``down path''). We only prove the properties for the up paths.
The proof for down paths can be obtained similarly by switching the role of
$\pred{provider}$ and $\pred{customer}$. 
Tuples $\pred{provider}$ and $\pred{customer}$ in SCION can be seen as 
counterparts of the $\pred{link}$ tuple in S-BGP, and 
tuple $\pred{beaconIni}$ and tuple $\pred{beaconFwd}$ correspond to tuple
$\pred{advertise}$. The definition of route
authenticity in SCION, denoted $\varphi_\textit{authS}$, is defined as:

\fbox{%
  \parbox{0.9\textwidth}{%
 $\varphi_\textit{authS}$ = $\forall n,m,t,\f{td},\f{itf},\f{tl},
\f{ol},\f{sl},\f{sg}, \\
~~~~~~~~~~~~~~ \pred{honest}(n) \conjunc \\
~~~~~~~~~~~~~~ (\pred{beaconIni}(@m,n,\f{td},\f{itf},\f{tl},\f{ol},\f{sl}
  ,\f{sg})@(n, t) \disj \\
~~~~~~~~~~~~~~ \pred{beaconFwd}(@m,n,\f{td},\f{itf},\f{tl},\f{ol},\f{sl}
  ,\f{sg})@(n, t)) \\
~~~~~~~~~~~~~~ \imp \pred{goodInfo}(\f{t},\f{td},\f{n},\f{sl},\f{itf})$
}
}\bigskip

\noindent Formula $\varphi_\textit{authS}$ asserts a property
$\pred{goodInfo}(t, \f{td}, n, \f{sl}, \f{itf})$ on any beacon tuple
generated by node $n$, which is either a TD core or an ordinary AD.

The definition of $\pred{goodInfo}$ is shown in Figure~\ref{fig:goodinfo}.  
Predicate $\pred{goodInfo}(t, \f{td}, \f{n},
\f{sl} ,\f{itf})$ takes five arguments: $t$ represents the time,
$\f{td}$ is the identity of the TD that the path lies in, $\f{n}$ is the
node that verifies the beacon containing the interface field $\f{itf}$, and
$\f{sl}$ is the signature list associated with the path.
$\pred{goodInfo}(t, \f{td}, \f{n}, \f{sl} ,\f{itf})$ makes sure that
each AD present in the interface field
$\f{itf}$ does have the specified links to its provider and customer
respectively. Also, for each non-core AD, there always exists a verified
beacon corresponding to the path from the TD core to it.

\begin{figure}[t!]
\begin{mathpar}
\mprset{flushleft}
\inferrule*{
 ~~~~~~~~~~~~~~~\pred{coreTD}(\f{ad},c,td,\f{ctf})@(\f{ad}, t)
\\
\pred{Honest}(c) \imp \exists t', t'\leq t \conjunc \pred{customer}(c,n,\f{ceg})@(c,t')
}{
  \pred{goodInfo}(t, \f{td}, \f{ad}, \f{nil}, c::\f{ceg}::n::\f{nil})}
\and
\mprset{flushleft}
\inferrule*{
\pred{coreTD}(\f{ad},c,td,\f{ctf})@(\f{ad},t)
\\\\
\pred{Honest}(n) \imp \exists t', t'\leq t \conjunc
\pred{provider}(n,c,\f{nig})@(n,t')\conjunc\pred{customer}(n,m,\f{neg})@(n,t') 
\\~~~~~~~~~~~~~~~~ \conjunc \exists td',tl,ol,sg,s,\pred{verifiedBeacon}(n,td',c::\f{ceg}::n::\f{nig}::nil,
 tl,
\\~~~~~~~~~~~~~~~~~~~~~~~~~~~~~~~~~~~~~~~~~~~~~~~~~~~~~~~~~~~~
 ol,s::\f{nil})@(n,t') 
\\\\
\pred{goodInfo}(t,td,\f{ad}, \f{nil}, (c::\f{ceg}::n::\f{nil}))
}{
\pred{goodInfo}(t, \f{td}, \f{ad}, s::\f{nil}, 
c::\f{ceg}::n::\f{nig}::\f{neg}::m::\f{nil})
}\and
\inferrule*{
\pred{Honest}(n) \imp \exists t', t'\leq t \conjunc \pred{provider}(n,h,\f{nig})@(n,t')
\conjunc\pred{customer}(n,m,\f{meg})@(n,t')
\\~~~~~~~~~~~~~~~~~ \conjunc\exists td', tl,ol,sg,s,sl,
\pred{verifiedBeacon}(n,td',p'{++} h::\f{hig}::\f{heg}::n::\f{nig},\\
\\~~~~~~~~~~~~~~~~~~~~~~~~~~~~~~~~~~~~~~~~~~~~~~~~~~~~~~~~~~
tl,ol,s::\f{sl})@(n,t').
\\
\pred{goodInfo}(t,td,\f{ad}, \f{sl}, p'++h::\f{hig}::\f{heg}::n::\f{nil})
}{
\pred{goodInfo}(t, \f{td}, n, s::\f{sl},
~~p'{++} h::\f{hig}::\f{heg}::n::\f{nig}::\f{neg}::m::\f{nil})
}
\end{mathpar}
\caption{Definitions of \pred{goodInfo}}
\label{fig:goodinfo}
\end{figure}

More concretely, the definition of $\pred{goodInfo}$ considers three cases. The base
case is when a TD core $c$ initializes an interface field
$c::\f{ceg}::n::\f{nig}::nil$ and sends it to AD $n$. We require that
$c$ be a TD core and $n$ be its customer. The next two cases are
similar, they both require the current AD $n$ have a link to its
preceding neighbor, represented by $\pred{provider}$, as well as one
to its downstream neighbor, represented by $\pred{customer}$. In
addition, a \pred{verifiedBeacon} tuple should exist, representing an
authenticated route stored in the database, with all inside signatures
properly verified. The difference between these two cases is caused by two
possible types of an AD's provider: TD core and non-TD core.

The proof strategy is exactly the same as that used in proof of $\pred{goodPath}$
about S-BGP. To prove $\varphi_\textit{authS}$, we first prove \scion
has a stronger program invariant $\varphi_{I}$:

\vspace{3pt}
\fbox{%
  \parbox{0.55\textwidth}{%
(b) $\cdot; \cdot\vdash \runby{\prog_\textit{sci}}{n}: \{i, y_b,
y_e\}.\varphi_{I}(i, y_b, y_e)$
}
}
\smallskip

\noindent where $\varphi_{I}(i, y_b, y_e)$ is defined as:

\vspace{3pt}
\fbox{%
  \parbox{0.9\textwidth}{%
$\varphi_{I}(i, y_b, y_e) = 
  \bigwedge_{p\in\headof{sci}}\forall t,\forall\vec{x}, y_b\leq t<y_e\conjunc
 p(\vec{x})@(i, t)
  \imp \varphi_{p}(i, t, \vec{x})$.
}
}
\bigskip

\noindent Especially, $\varphi_{p}$ for \pred{beaconIni} and \pred{beaconFwd} are as follows:

\vspace{3pt}
\fbox{%
  \parbox{0.85\textwidth}{%
$\varphi_\textit{beaconIni}(i,t,m,n,\f{td},\f{itf},\f{tl},\f{ol},\f{sl},\f{sg})=\pred{goodInfo}(t,\f{td},n,\f{sl},\f{itf})$  \\
\indent$\varphi_\textit{beaconFwd}(i,t,m,n,\f{td},\f{itf},\f{tl},\f{ol},\f{sl},\f{sg})=\pred{goodInfo}(t,\f{td},n,\f{sl},\f{itf})$
}
}
\bigskip

\noindent As in S-BGP, (b) can be proved using \rulename{Inv} rule, whose premises are verified in Coq. 
After (b) is proved, we can deduce 
$\varphi'=\forall n, t', \pred{Honest}(n) \imp \varphi_I(n, t', -\infty)$ by
applying \rulename{Honest} rule to (b). Finally, $\varphi_\textit{authS}$ is
proved by showing that $\varphi' \imp \varphi_\textit{authS}$, which is straightforward.

At the end of the path construction phase, an AD needs to upload its selected paths
to the TD core for future queries (i.e. rules $pc1-pc4$ in
Table~\ref{tab:scion}).
The uploading process uses the forwarding mechanism in SCION, which provides
hop-by-hop authentication. An AD who wants to send traffic to another AD 
attaches each data packet with the opaque field extracted from a beacon 
received during the path construction phase. The opaque field contains MACs of the 
ingress and egress of all ADs on the intended path. When the data packet is sent
along the path, each AD en-route re-computes the MAC of intended ingress and
egress using its own private key. This MAC is compared with the one
contained in the opaque field. If they are the
same, the AD knows that it has agreed to receiving/sending
packets from/to its neighbors during path construction phase and forwards the
packet further along the path. Otherwise, it drops the
data packet. 

The formal definition of data path authenticity in SCION is defined as:

\limin{fix indentation\\}
\vspace{3pt}
\fbox{%
  \parbox{0.93\textwidth}{%
 $\varphi_\textit{authD}$ = $\forall m,n,t,\f{src},\f{core},\f{itf},
\f{opqD},\f{opqU},\f{pt},\\
~~~~~~~~~~~~~~\pred{honest}(n) \conjunc \\
~~~~~~~~~~~~~~\pred{pathUpload}(@m,n,\f{src},\f{core},\f{itf},\f{opqD},\f{opqU}
  ,\f{pt})@(n, t) \imp \\
~~~~~~~~~~~~~~~~ \pred{goodFwdPath}(\f{t},\f{n},\f{opqU},\f{pt})$
}
}
\bigskip

\noindent Formula $\varphi_\textit{authD}$ asserts property $\pred{goodFwdPath}
(\f{t},\f{n},\f{opqU},\f{pt})$ on any tuple $\pred{pathUpload}$ sent by
a customer AD to its provider. 
There are four arguments in $\pred{goodFwdPath}(\f{t},\f{n},\f{opqU},\f{pt})$:
$\f{t}$ is the time. $\f{n}$ is the node
who sent out $\pred{pathUpload}$ tuple. $\f{opqU}$ is a list of opaque fields
for forwarding. $\f{pt}$ is a pointer to $\f{opqU}$, indicating the next opaque
field to be checked. Except time
$\f{t}$, all arguments in $\pred{goodFwdPath}(\f{t},\f{n},\f{opqU},\f{pt})$
are the same as those in $\pred{pathUpload}$, whose arguments are
described in Figure~\ref{fig:scion:tuples}.
$\pred{goodFwdPath}(\f{t},\f{n},\f{opqU},\f{pt})$ states that whenever an
AD receives a packet, it has direct links to its
provider and customer as indicated by the opaque field in the packet. In
addition, it must have verified a beacon with a path containing this neighboring
relationship. 

\begin{figure}[t!]
\begin{mathpar}
\mprset{flushleft}
\inferrule*{
 pt = 0}{
  \pred{goodFwdPath}(\f{t},\f{n},\f{opqU},\f{pt})}
\and

\mprset{flushleft}
\inferrule*{
\pred{Honest}(n) \imp \exists t',m,td,ctf, t'\leq t \\\\
~~~~~~~~~~~~~~~~\conjunc \pred{coreTD}(\f{n},n,td,\f{ctf})@(\f{n},t')
\\\\
~~~~~~~~~~~~~~~~\conjunc \pred{customer}(n,m,\f{ceg})@(n,t')
}{
\pred{goodFwdPath}(\f{t},\f{n},opq'{++}[\f{ceg}::\f{mac}::\f{nil}]::\f{nil},\\
~~~~~~~~~~~~~~~~~~length(opq'{++}[\f{ceg}::\f{mac}::\f{nil}]::\f{nil}))
}\and
\inferrule*{
0 < pt \conjunc pt < length(opq'{++}[\f{nig}::\f{neg}::\f{mac'}::\f{nil}]::[\f{ceg}::\f{mac}::\f{nil}]::nil\conjunc \\
\pred{Honest}(n) \imp \exists t',h,m, t'\leq t \\\\
~~~~~~~~~~~~~~~~~~\conjunc \pred{provider}(n,h,\f{nig})@(n,t') \\\\
~~~~~~~~~~~~~~~~~~\conjunc\pred{customer}(n,m,\f{meg})@(n,t') \\\\
~~~~~~~~~~~~~~~~~~ \conjunc\exists td', tl,sg,sl,
\pred{verifiedBeacon}(n,td',h::\f{ceg}::n::\f{nig},\\
\\~~~~~~~~~~~~~~~~~~~~~~~~~~~~~~~~~~~~~~~~~~~~~~~~~~~~~~~~~~
tl,[\f{ceg}::\f{mac}::nil]::nil,\f{sl},\f{sg})@(n,t')
}{
\pred{goodFwdPath}(\f{t},\f{n},opq'{++}[\f{nig}::\f{neg}::\f{mac'}::\f{nil}]::[\f{ceg}::\f{mac}::\f{nil}]::\f{nil},\f{pt})}\and
\inferrule*{
0 < pt \conjunc \\
pt < length(opq'{++}[\f{nig}::\f{neg}::\f{mac'}::\f{nil}]::[\f{hig}::\f{heg}::\f{mac}::\f{nil}]{++}\f{opq''}) \conjunc \\
\pred{Honest}(n) \imp \exists t',h,m, t'\leq t \\\\
~~~~~~~~~~~~~~~~~~\conjunc \pred{provider}(n,h,\f{nig})@(n,t') \\\\
~~~~~~~~~~~~~~~~~~\conjunc\pred{customer}(n,m,\f{meg})@(n,t') \\\\
~~~~~~~~~~~~~~~~~~ \conjunc\exists td', tl,sg,sl,p',p'',
\pred{verifiedBeacon}(n,td',p'{++}h::\f{hig}::\f{heg}::n::\f{nig},\\
\\~~~~~~~~~~~~~~~~~~~~~~~~~~~~~~~~~~~~~~~~~~~~~
tl,p''{++}[\f{hig}::\f{heg}::\f{mac}::nil]::nil,\f{sl},\f{sg})@(n,t')
}{
\pred{goodFwdPath}(\f{t},\f{n},opq'{++}[\f{nig}::\f{neg}::\f{mac'}::\f{nil}]::[\f{hig}::\f{heg}::\f{mac}::\f{nil}]{++}\f{opq''},\f{pt})}
\end{mathpar}
\vspace{-20pt}
\caption{Definitions of \pred{goodFwdPath}}
\label{fig:goodfwd}
\end{figure}

The definition of $\pred{goodFwdPath}(\f{t},\f{n},\f{opqU},\f{pt})$ is given in
Figure~\ref{fig:goodfwd}. There are four cases. 
The base
case is when $\f{pt}$ is 0, which means nothing has been verified. In this case $\pred{goodFwdPath}$
holds trivially. If $\f{pt}$ is equal to the length of opaque field list,
meaning all opaque fields have been verified already, then based on SCION
specification, the last opaque field should be that of the TD core. Being a
TD core requires a certificate ($\pred{coreTD}$), and a neighbor customer along the
path ($\pred{customer}$). When $\f{pt}$ does not point to the head or the tail
of the opaque field list, node $\f{n}$ should have a neighbor provider($\pred{provider}$), and a neighbor customer($\pred{customer}$). It must also have
received and processed a $\pred{verifiedBeacon}$ during path construction.
The second and third cases both cover this scenario. The second case applies when a
node $\f{n}$'s provider is TD core, while in the third case, $\f{n}$'s provider and
customer are both ordinary TDs.

SCION uses MAC for integrity check during data forwarding, so we use
the following axiom about MAC.
It states that if a node $\f{n}$ verifies a MAC, using $\f{n'}$'s key $\f{k}$, there
must have been a node $\f{n''}$ who created the MAC at an earlier time $\f{t'}$.

\vspace{3pt}
\fbox{%
  \parbox{0.9\textwidth}{%
$A_\textit{mac}$ = $\forall msg,m,k,n,n',t,\\
~~~~~~~~~~~~\pred{verifyMac}(msg, m, k)@(n, t) \conjunc 
\pred{privateKeys}(n, n', k)@(n, t) \conjunc \\
~~~~~~~~~~~~ \exists n'', \pred{Honest}(n'') \imp\exists t', t'< t\conjunc
\pred{privateKeys}(n'', n', k)@(n'', t) \conjunc \\
~~~~~~~~~~~~~~~~~~~~~~~~~~~~~~~~~~~~~~~~~\pred{mac}(n'', msg, m)@(n'', t')$
}
}
\bigskip

\noindent In SCION, each node should not share its own private key with other nodes.
This means, for each specific MAC, only the node who generated it can verify
its validity. This fact simplifies the axiom:

\vspace{3pt}
\fbox{%
  \parbox{0.8\textwidth}{%
$A'_\textit{mac}$ = $\forall msg,m,k,n,t, \\
~~~~~~~~~~~~\pred{verifyMac}(msg, m, k)@(n, t) \conjunc 
\pred{privateKeys}(n, k)@(n, t) \conjunc \\
~~~~~~~~~~~~\pred{Honest}(n) \imp\exists t', t'< t\conjunc \pred{mac}(n, msg, m)@(n, t')$
}
}
\bigskip

\noindent The rest of the proof follows the same strategy as that of
$\pred{goodPath}$ and $\pred{goodInfo}$. Interested readers can refer to our
proof online for details.

\subsection{Comparison between S-BGP and SCION}
\label{sec:casestudy:comparison}
In this section, we compare the difference
between the security guarantees provided by S-BGP and SCION. In terms of practical
route authenticity, there is little difference between
what S-BGP and SCION can offer. This is not surprising, as the kind of
information that S-BGP and SCION sign at path construction phase is
very similar.

Though both use layered-signature to protect the routing information,
signatures in S-BGP are not technically layered---ASes in S-BGP only
sign the path information, not including previous signatures. On the
other hand, ADs in SCION sign the previous signature so signatures in
SCION are nested.

Consider an AS $n$ in S-BGP that signed the path $p$ twice,
generating two signatures: $s$ and $s'$. An attacker, upon receiving a
sequence of signatures containing $s$, can replace $s$ with $s'$
without being detected. This attack is not possible in SCION, as
attackers cannot extract signatures from a nested signature. 

SCION also provides stronger security guarantees than S-BGP in data
forwarding.
Though S-BGP does not explicitly state the process of data forwarding,
we can still compare its IP-based forwarding to SCION's forwarding
mechanism. Like BGP, an AS running S-BGP maintains a routing table
on all BGP speaker routers that connect to peers in other
domains. The routing table is an ordered collection of forwarding
entries, each represented as a pair of $\langle$IP prefix, next
hop$\rangle$. Upon receiving a packet, the speaker searches its routing table
for IP prefix that matches the destination IP address in the IP
header of the packet, and forwards the packet on the port
corresponding to the next hop based on table look-up.
This next hop must have been authenticated, because only after an
S-BGP update message has been properly verified will the AS insert
the next hop into the forwarding table.

However, SCION provides stronger security guarantee over S-BGP in
terms of the last hop of the packet. An AS $\f{n}$ running S-BGP has no way
of detecting whether a received packet is from legitimate neighbor
ASes who are authorized to forward packets to $\f{n}$. Imagine that $\f{n}$ has
two neighbor ASes, $\f{m}$ and $\f{m'}$. $\f{n}$ knows a
route to an IP prefix $\f{p}$ and is only willing to advertise the
route to $\f{m}$.  Ideally, any packet from $\f{m'}$ through $\f{n}$
to $\f{p}$ should be rejected by $\f{n}$. However, this may not happen
in practice for AS's who run S-BGP for routing. 
As long as its IP destination is $\f{p}$, a packet will be forwarded by $\f{n}$,
regardless of whether it is from $\f{m}$ or $\f{m'}$.
On the other hand, SCION routers are able to 
discard such packets by verifying the MAC in the opaque field, since $\f{m}$
cannot forge the MAC information embedded in the beacon.

\subsection{Empirical evaluation}
\label{sec:casestudy:evaluation}
We use RapidNet~\cite{rapidneturl} to generate low-level
implementation 
of S-BGP and SCION from \Langsec encoding. We validate the low-level
implementation 
in the ns-3 simulator~\cite{ns3}. 
Our experiments are performed on a synthetically generated topology 
consisting of $40$ nodes, where each node runs the generated 
implementation of the \Langsec program. 
The observed execution traces and communication patterns
match the expected protocol behavior.

\section{Related Work}
\noindent{\bf Cryptographic Protocol Analysis.} The analysis of
cryptographic
protocols~\cite{datta07:entcs,datta08:secrecy,changhua05:tls,Paulson:97,maude-npa,proverif-journal,murphi:dnssec,Garg:2010} has been an active area of research.  Compared with
cryptographic protocols, secure routing protocols have to deal with
arbitrary network topologies and the programs of the protocols are
more complicated: they may access local storage and commonly include
recursive computations. Most model-checking techniques are ineffective
in the presence of those complications.

\noindent{\bf Verification of Trace Properties.} A closely related
body of work is logic for verifying trace properties of programs
(protocols) that run concurrently with
adversaries~\cite{datta07:entcs,Garg:2010}.  We are inspired by their
program logic that requires the asserted properties of a program to
hold even when that program runs concurrently with adversarial
programs. One of our contributions is a general program logic for a
declarative language \Langsec, which differs significantly from an
ordinary imperative language. The program logic and semantics
developed here apply to other declarative languages that use bottom-up
evaluation strategy.

\noindent{\bf Networking Protocol Verification.} Recently, several papers
have investigated the verification of route authenticity properties on
specific wireless routing protocols for mobile networks
~\cite{acd:csf10,acd:cade11,four-node:post12}. 
They have showed that identifying attacks on route authenticity can be
reduced to constraint solving, and that the security
analysis of a specific route authenticity property that depends on the
topologies of network instances can be reduced to checking these
properties on several four-node topologies. In our own prior
work~\cite{chen:wripe12}, we have verified route authenticity
properties on variants of S-BGP using a combination of manual proofs
and an automated tool, Proverif~\cite{proverif-tool}.  The modeling and
analysis in these works are specific to the protocols and the route
authenticity properties.  Some of the properties that we
verify in our case study are similar.  However, we propose a
general framework for leveraging a declarative programming language
for verification and empirical evaluation of routing protocols. The
program logic proposed here can be used to verify generic safety
properties of \Langsec programs.\reviewer{are there any other
interesting generic properties of SeNDLog programs you have in mind?
otherwise this is a very vague claim.}

There has been a large body of
work on verifying the correctness of various network protocol design
and implementations using proof-based and model-checking
techniques~\cite{dvverification,maude,cmcnets}.  The program logic
presented here is customized to proving safety properties of \Langsec
programs, and may not be expressive enough to verify complex
correctness properties.  However, the operational semantics for
\Langsec can be used as the semantic model
for verifying protocols encoded in \Langsec using other techniques.

\reviewer{it would help to mention more explicitly what kind of features are challenging for automatically discharging VCs...things like time, location, inductive
relations...
Another speculative thing you could discuss is how your framework
would handle secrecy/encryption. It seems at the moment very much
tailored towards authenticity.}

\section{Conclusion and Future Work}

We have designed a program logic for verifying secure routing
protocols specified in the declarative language \Langsec. 
We have integrated verification into
a unified framework  for formal analysis and empirical evaluation of secure routing
protocols. As future
work, we plan to expand our use cases, for example, to investigate
mechanisms for securing the data (packet forwarding)
plane~\cite{icing}. In addition, as an alternative to Coq, we are also
exploring the use of automated first-order logic theorem provers to
automate our proofs.

\bibliographystyle{plain}
\bibliography{paper}

\clearpage
\appendix
\section{Proof of Theorem~\ref{thm:soundness}}
\label{app:soundness}
{\small
By mutual induction on the derivation $\ee$. 
The rules for standard first-order logic inference rules 
 $\Sigma;\Gamma\vdash\varphi$ are
straightforward. We show the case when \ee ends in the \rulename{Honest} rule.

\vspace{-5pt}
\begin{description}
\item[Case:] The last step of \ee is \rulename{Honest}.
\begin{tabbing}
  \ee = 
\begin{mathpar}
\small
\inferrule*[right=Honest]{
 \ee_1::\Sigma; \Gamma \vdash \runby{\prog}{i} : \{i, y_b,y_e\}.\varphi(i, y_b,y_e)
\\ \ee_2::\Sigma; \Gamma \vdash \pred{honest}(\nodeid, \prog(\nodeid), t)
}{
\Sigma;\Gamma \vdash \forall t', t'>t, \varphi(\nodeid, t, t')
}
\end{mathpar} \\
Given $\sigma$, $\trace$ s.t. $\trace\vDash \Gamma\sigma$, by I.H. on
$\ee_1$ and $\ee_2$
\\~~ \= (1) $\Gamma\sigma\vDash \runby{(\prog)\sigma}{i} :  \{i, y_b,y_e\}.(\varphi(i,
y_b,y_e))\sigma$ 
\\\> (2) $\trace\vDash (\pred{honest}(\nodeid, \prog(\nodeid), t))\sigma$
\\ By (2), 
\\\> (3) at time $t\sigma$, $\nodeid\sigma$ starts to run program
($(\prog)\sigma$)
\\ By (1) and (3), given any $T$ s.t. $T>t\sigma$
\\\> (4) $\trace\vDash \varphi\sigma(\nodeid\sigma, t\sigma, T)$
\\ Therefore,
\\\> (5) $\trace\vDash  (\forall t', t'>t, \varphi(\nodeid, t, t'))\sigma$
\end{tabbing}
\vspace{-5pt}
\item[Case:] \ee ends in \rulename{Inv} rule.
\\ Given $\trace$, $\sigma$
  such that $\trace\vDash \Gamma\sigma$, and at time $\tau_b$, 
 node \nodeid's local state is (\nodeid, [], [], $\prog(\nodeid)$),
 given any time point $\tau_e$ such that  $\tau_e \geq \tau_b$,
\\ let $\varphi = 
 (\bigwedge_{p\in\headof{\prog}}\forall t,\forall\vec{x}, \tau_b\leq t<\tau_e\conjunc
 p(\vec{x})@(\nodeid, t)
  \imp \varphi_{p}(\nodeid, t, \vec{x}))\sigma$
\\ we need to show $\trace \vDash \varphi$
\\ By induction on the length of $\trace$
\item[subcase:] $|\trace| = 0$, $\trace$ has one state and is of the form $\steps{\tau}\conf$
\\ By assumption (\nodeid, [], [], $[\prog]_{\nodeid}$) $\in\conf$
\\ Because the update list is empty, $\nexists \sigma_1$, s.t. 
$\trace\vDash (p(\vec{x})@(\nodeid, t)) \sigma\sigma_1$
\\ Therefore, $\trace \vDash \varphi$ trivially.
\item[subcase:] $\trace = \trace' \steps{\tau} \conf$ 
\\We examine all possible steps allowed by the operational semantics.
\\To show the conjunction holds, we show all clauses in the conjunction
are true by construct a generic proof for one clause.
\begin{description}
\item[case:] \rulename{DeQueue} is the last step.
\begin{tabbing}
 Given a substitution $\sigma_1$ for $t$ and $\vec{x}$ s.t. 
 $\trace\vDash (\tau_b\leq t<\tau_e\conjunc p(\vec{x})@(\nodeid, t))
 \sigma\sigma_1$
\\By the definitions of semantics, and \rulename{DeQueue} merely moves
messages around
\\~~\= (1) $(p(\vec{x}))\sigma\sigma_1$ is on trace $\trace'$
\\\> (2) $\trace'\vDash (\tau_b\leq t<\tau_e\conjunc p(\vec{x})@(\nodeid, t))
 \sigma\sigma_1$
\\By I.H. on $\trace'$,
\\\> (3) $\trace'\vDash (\forall t,\forall\vec{x}, \tau_b\leq t<\tau_e\conjunc
 p(\vec{x})@(\nodeid, t)
  \imp \varphi_{p}(\nodeid, t, \vec{x}))\sigma$
\\ By (2) and (3) 
\\\> (4) $\trace'\vDash \varphi_{p}(\nodeid, t, \vec{x})\sigma\sigma_1$
\\ By $\varphi_{p}$ is closed under trace extension and (4),
\\\>  $\trace\vDash \varphi_{p}(\nodeid, t, \vec{x})\sigma\sigma_1$
\\ Therefore, $\trace\vDash \varphi$ by taking the conjunction of all
the results for such $p$'s.
\end{tabbing}
\item[case:] \rulename{NodeStep} is the last step. Similar to the
  previous case, we examine every tuple $p$ generated by \prog to show
  $\trace\vDash\varphi$. When $p$ was generated on $\trace'$, the
  proof proceeds in the same way as the previous case. 
  We focus on the cases where $p$ is generated in the last step.
\\We need to show that $\trace\vDash 
(\forall t,\forall\vec{x}, \tau_b\leq t<\tau_e\conjunc p(\vec{x})@(\nodeid, t)
  \imp \varphi_{p}(\nodeid, t, \vec{x}))\sigma$
\\Assume the newly generated tuple is $(p(\vec{x})@(\nodeid,
\tau_p))\sigma\sigma_1$, where $\tau_p \geq \tau$
\\We need to show that  $\trace\vDash 
(\varphi_{p}(\nodeid, \tau_p, \vec{x}))\sigma\sigma_1$
\begin{description}
\item[subcase:] \rulename{Init} is used
\begin{tabbing}
In this case, only rules with an empty body are fired ($r =h(\vec{v})
\derives .$). 
\\ By expanding the last premise of the \rulename{Inv} rule, and
$\vec{v}$ are all ground terms,
\\~~\= (1) $\ee_1::\Sigma;\Gamma \vdash  \forall i,\forall t,
\varphi_h(i, t, \vec{v})$
\\By I.H. on $\ee_1$ (here $\ee_1$ is a smaller derivation than $\ee$,
so we can directly invoke 1)
\\\> (2)  $\trace\vDash (\forall i, \forall t,
\varphi_h(i, t, \vec{v}))\sigma$ 
\\ By (2)
\\\> $\trace\vDash (\varphi_h(\nodeid, \tau_p, \vec{y}))\sigma\sigma_1$
\end{tabbing}
\item[subcase:] \rulename{RuleFire} is used. 
\\We show one case where $p$ is not an aggregate and one where $p$ is.
\item[subsubcase:] \rulename{InsNew} is fired
\begin{tabbing}
By examine the $\Delta r$ rule,
\\ ~~\= (1) exists $\sigma_0\in \substof(\tuples^\nu, \tuples, \rules,
k, \vec{s})$ such
that $(p(\vec{x})@(\nodeid, t))\sigma\sigma_1 = (p(\vec{v})@(\nodeid, t))\sigma_0$
\\\>(2) for tuples ($p_j$) that are derived by node \nodeid, 
 $(p_j(\vec{s_j}))\sigma_0 \in \tuples^\nu$ or 
 $(p_j(\vec{s_j}))\sigma_0 \in \tuples$ 
 \\ By operational semantics, $p_j$ must have been generated on $\trace'$
\\\> (3)
 $\trace'\vDash(p_j(\vec{s}_j)@(\nodeid, \tau_p))\sigma_0$ 
\\ By I.H. on $\trace'$ and (3), the invariant for $p_j$ holds on
$\trace'$ 
\\\> (4) $\trace'\vDash(\varphi_{p_j}(\nodeid, \tau_p,
\vec{s}_j))\sigma_0$
\\ By $\varphi_p$ is closed under trace extension
\\\>(5) $\trace\vDash 
(p_j(\vec{x}_j)@(\nodeid, \tau_p)\conjunc\varphi_{p_j}(\nodeid, \tau_p,
\vec{s}_j))\sigma_0$
\\ For tuples ($q_j$) that are received by node \nodeid, using similar
reasoning as above
\\\> (6) $\trace\vDash (\pred{recv}(i, \pred{tp}(q_j, \vec{s}_j))@ \tau_p)\sigma_0$
\\ (7) For constraints ($a_j$), $\trace \vDash a_j\sigma_0 $
\\By I.H. on the last premise in \rulename{Inv} and (5) (6) (7)
\\\> (8) $\trace\vDash (\varphi_p(i, \tau_p, \vec{v}))\sigma_0$
\\ By (1) and (8),
$\trace\vDash (\varphi_p(i, \tau_p, \vec{y}))\sigma\sigma_1$
\end{tabbing}
\item[subsubcase:]\rulename{InsAggNew} is fired. 
\begin{tabbing}
When $p$ is an aggregated predicate, we additionally prove that
\\ every aggregate candidate predicate $p_\textit{agg}$ has the same invariant
as $p$. 
\\ That is (1) $\trace\vDash (\forall t,\forall\vec{x}, \tau_b\leq t<\tau_e\conjunc
 p_\textit{agg}(\vec{x})@(\nodeid, t)
  \imp \varphi_{p}(\nodeid, t, \vec{x}))\sigma$
\\The reasoning is the same as the previous case. 
\\We additionally show that (1) is true on the newly generated $p_\textit{agg}(\vec{t})$.
\end{tabbing}
\end{description}
\end{description}
\end{description}
}

\end{document}

%% file: macros.tex
\newcommand{\Langsec}{{SANDLog}\xspace}
\newcommand{\Lang}{{NDLog}\xspace}

\newcommand{\VCG}{{VCGen}\xspace}
\newcommand{\Paragraph}[1]{\vspace{3pt}\noindent{\bf #1}}

\newcommand{\bnfdef}{::=}
\newcommand{\bnfalt}{\,|\,}

\newcommand{\fv}{{\sf fv}}

\newcommand{\rulename}[1]{{\small\sc #1}}
\newcommand{\f}[1]{\mathit{#1}}
\newcommand{\figspace}{\vspace{5pt}}

\mathchardef\mhyphen="2D
\newcommand{\pred}[1]{\textsf{\small{#1}}\xspace}
\newcommand{\imp}{\mathrel{\supset}}
\newcommand{\conjunc}{\mathrel{\wedge}}
\newcommand{\disj}{\mathrel{\vee}}

\newcommand{\alist}{\textit{ags}}
\newcommand{\balist}{\textit{agB}}
\newcommand{\halist}{\textit{agH}}
\newcommand{\agf}{\ensuremath{F_\textit{agr}}}
\newcommand{\rbody}{\textit{body}}
\newcommand{\rules}{\textit{r}\xspace}
\newcommand{\predicate}{\textit{pred}\xspace}
\newcommand{\prog}{\textit{prog}\xspace}
\newcommand{\derives}{\ensuremath{\mathrel{:\!\!-}}}
\newcommand{\bop}{\ensuremath{\mathrel{\textit{bop}}}}

\newcommand{\sbgp}{\ensuremath{\prog_\textit{sbgp}}\xspace}
\newcommand{\scion}{\ensuremath{\prog_\textit{scion}}\xspace}

\newcommand{\trace}{\ensuremath{\mathcal{T}}\xspace}
\newcommand{\queue}{\ensuremath{\mathcal{Q}}\xspace}
\newcommand{\conf}{\ensuremath{\mathcal{C}}\xspace}

\newcommand{\ttrue}{\mathsf{true}}
\newcommand{\tfalse}{\mathsf{false}}
\newcommand{\nodeid}{\ensuremath{\iota}\xspace}

\newcommand{\lstate}{\ensuremath{\mathcal{S}}\xspace}
\newcommand{\tuples}{\ensuremath{\Psi}\xspace}
\newcommand{\updates}{\ensuremath{\mathcal{U}}\xspace}
\newcommand{\upd}{\textit{u}\xspace}
\newcommand{\del}{-\xspace}
\newcommand{\ins}{+\xspace}

\newcommand{\steps}[1]{\xrightarrow{#1}}
\newcommand{\stepsone}{\hookrightarrow}
\newcommand{\firesingle}{\textit{fireSingleR}\xspace}
\newcommand{\firerules}{\textit{fireRules}}

\newcommand{\runby}[2]{#1(#2)}
\newcommand{\headof}[1]{\textit{hdOf}(#1)}
\newcommand{\ruleof}[1]{\textit{rlOf}(#1)}
\newcommand{\baseof}[1]{\textit{BaseOf}(#1)}
\newcommand{\defeq}{\mathrel{\triangleq}}
\newcommand{\substof}{\rho}
\newcommand{\select}{\textit{sel}\xspace}
\newcommand{\genupd}{\textit{genUpd}\xspace}
\newcommand{\aggof}{\textit{Agg}}

\newcommand{\ee}{\ensuremath{\mathcal{E}}\xspace}

\newenvironment{packedenumerate}{\vspace{-3pt}
\begin{enumerate}
\setlength{\topsep}{0pt}
\setlength{\itemsep}{0pt}
\setlength{\partopsep}{0pt}
}{\end{enumerate}}

\newenvironment{packeditemize}{\vspace{-3pt}
\begin{itemize}
\setlength{\topsep}{0pt}
\setlength{\itemsep}{0pt}
\setlength{\partopsep}{0pt}
}{\end{itemize}}

\newenvironment{NDlog}{\vspace{-1.5mm}\begin{alltt}\footnotesize}{\end{alltt}\vspace{-1.5mm}}

 \newcommand{\notes}[1]{}
\newcommand{\limin}[1]{\notes{Limin says: #1}}
\newcommand{\chen}[1]{\notes{Chen says: #1}}

\newcommand{\reviewer}[1]{}

\algnewcommand{\LineComment}[1]{\State \(\triangleright\) #1}
\algnewcommand\algorithmicinsert{\textbf{Insert }}
\algnewcommand\algorithmicwhere{\textbf{Where }}
\algnewcommand\algorithmicoutput{\textbf{Output }}
\algnewcommand\algorithmiccase{\textbf{Case }}
\algtext*{EndIf}
\algtext*{EndFor}